%% file: ANA-HIGP-2024-21-PAPER.tex
\documentclass[PAPER, atlasdraft=false,cernpreprint, USenglish, texmf, orcidlogo]{atlasdoc}
\usepackage{amsmath}
\usepackage{atlaspackage}
\usepackage{tabularx}
\usepackage{multirow}
\usepackage{atlasbiblatex}

\usepackage{atlasphysics}
\graphicspath{{logos/}{figures/}}

\usepackage{ANA-HIGP-2024-21-PAPER-defs}

\AtlasTitle{Search for $H\rightarrow c\bar{c}$ and measurement of $H \rightarrow b\bar{b}$ in vector-boson fusion production with the ATLAS detector}
\AtlasAbstract{A search for Standard Model (SM) Higgs bosons produced via vector-boson fusion at the Large Hadron Collider and decaying into a charm quark--antiquark pair (\Hcc) is presented. The datasets used correspond to integrated luminosities of $37.5$ and $51.5~\invfb$ and were collected by the ATLAS detector from proton--proton collisions at $\sqrt{s}=13$ and $13.6~\TeV$ respectively. The observed (expected) upper limit on the \Hcc\ production cross-section times branching ratio is $41~(28)$ times the SM prediction at 95\% confidence level. Combining this search with the previous \Hcc\ search in associated production with a $W$ or $Z$ boson yields an observed (expected) limit on the Higgs--charm Yukawa coupling modifier of $|\kappa_c|<4.7~(3.9)$. Higgs bosons decaying into a bottom quark--antiquark pair (\Hbb) is measured simultaneously using the $51.5~\invfb$ dataset at $\sqrt{s}=13.6~\TeV$, with an observed signal strength of $0.97^{+0.57}_{-0.50}$ relative to the SM expectation. When combined with previous \Hbb\ results at 13~TeV, the observed (expected) significance reaches 3.2 (3.6) standard deviations, providing evidence for \Hbb\ events from vector-boson fusion.}

\AtlasRefCode{HIGP-2024-21}

\PreprintIdNumber{CERN-EP-2025-247}

\AtlasJournal{Phys.\ Rev.\ Lett.}
\AtlasCoverEgroupAnalysisTeam{atlas--analysis-team@cern.ch}

\hypersetup{pdftitle={ATLAS document},pdfauthor={The ATLAS Collaboration}}

\begin{document}

\maketitle

\label{sec:intro}

The discovery of the Higgs boson in 2012 at the Large Hadron Collider (LHC)~\cite{CMS-HIG-12-028,HIGG-2012-27} was a landmark achievement in particle physics, confirming the existence of the last missing piece of the Standard Model (SM). Since the discovery, properties of the Higgs boson have been studied extensively. These include its mass~\cite{HIGG-2013-12,HIGG-2014-14,HIGG-2016-33,HIGG-2019-16,HIGG-2020-07,CMS-HIG-19-004,CMS-HIG-21-019-own}, spin~\cite{CMS-HIG-12-041,CMS-HIG-14-018,HIGG-2013-01,HIGG-2013-14,HIGG-2013-17,HIGG-2022-04}, and couplings to gauge bosons~\cite{HIGG-2016-22, HIGG-2016-25, HIGG-2017-11,CMS-HIG-13-002,CMS-HIG-14-028,CMS-HIG-18-002,CMS-HIG-13-023,CMS-HIG-19-002,CMS-HIG-22-008,HIGG-2013-13,HIGG-2013-15} and third-generation fermions~\cite{HIGG-2018-04,CMS-HIG-16-044,CMS-HIG-18-016, CMS-HIG-16-043,HIGG-2013-32}, which are all found to align with the SM predictions. More recently, both ATLAS and CMS have reported evidence for the coupling of the Higgs boson to muons~\cite{HIGP-2024-11,CMS-HIG-19-006}, providing the first direct probe of Higgs boson interactions with second-generation fermions. However, in the quark sector, one of the key properties that remains to be explored is the coupling of the Higgs boson to charm quarks.

The search for Higgs boson decay into a charm quark--antiquark pair (\Hcc) remains challenging because of the Higgs boson's smaller Yukawa coupling to charm quarks than to bottom quarks, the overwhelming quantum chromodynamics (QCD) multijet backgrounds, and the challenges in experimentally distinguishing charm jets from light-flavor or bottom jets. Both the ATLAS~\cite{HIGG-2020-20} and CMS~\cite{CMS-HIG-21-008} collaborations have constrained the Higgs--charm Yukawa coupling through searches for \Hcc\ in associated production with a vector boson (\VH, $V=W,Z$). In addition, the CMS Collaboration~\cite{CMS-HIG-24-018} has also performed a search in associated production with a top-quark pair (\ttH). The most stringent observed (expected) limits are $11.5~(10.6)$ times the SM prediction, from ATLAS in the \VH\ production mode, and $9.3~(5.6)$ times the SM prediction, from CMS using a combination of \VH\ and \ttH\ production modes.

This Letter presents a search for \Hcc\ by exploiting the second-largest Higgs boson production mode, vector-boson fusion (VBF). In parallel, the analysis also targets \Hbb\ decays in VBF production. The bottom-quark Yukawa coupling plays a central role in the Higgs sector, and while \Hbb\ has been measured in associated-production modes (\VH and \ttH\ )~\cite{HIGG-2020-20,HIGG-2020-24, CMS-HIG-18-016, CMS-HIG-19-011}, it has not yet been observed in the VBF production mode. %
Measuring the VBF production cross-section in the \Hbb\ decay mode is also important for Higgs boson measurements within kinematic fiducial regions~\cite{deFlorian:2016spz}, particularly at high transverse momentum~\cite{HIGG-2019-04}.

The VBF production mode is characterized by the presence of two forward jets with large invariant mass and rapidity separation, produced via the exchange of a virtual vector boson between the two initial-state quarks. A new inclusive trigger exploiting this topology~\cite{TRIG-2022-01} was deployed in ATLAS partway through 2018, during \RunTwo of the LHC. It achieves an efficiency for \vbfhbb\ events  comparable to that of the $b$-tagging-based trigger used in the Run-2 \vbfhbb\ analysis~\cite{HIGG-2019-04}, while also enabling the first search for the \vbfhcc\ process. The \Hcc\ search utilizes proton--proton collision data collected by the ATLAS detector at two center-of-mass energies: $37.5~\invfb$ at $\sqrt{s}=13$~TeV from part of the 2018 Run-2 period and $51.5~\invfb$ at $\sqrt{s}=13.6$~TeV during 2022--2023 in \RunThr, yielding a total integrated luminosity of $89.0~\invfb$. The \Hbb\ process is measured only with the Run-3 data, allowing direct combination with the analysis performed on full Run-2 dataset~\cite{HIGG-2019-04}. The analysis strategy follows Ref.~\cite{HIGG-2019-04}, with improvements in multivariate techniques, the adoption of the inclusive VBF trigger, and the use of a novel transformer-based flavor-tagging algorithm~\cite{FTAG-2023-05} for jet flavor identification.


%
\label{sec:detector}

The ATLAS detector~\cite{PERF-2007-01} is a multipurpose particle detector designed to investigate a wide range of physics processes at the LHC. It consists of an inner tracking detector surrounded by a superconducting solenoid, electromagnetic and hadronic calorimeters, and a muon spectrometer. Events are triggered by a two-level trigger system, with the first level implemented in hardware and the second level in software~\cite{TRIG-2019-04, TRIG-2022-01}. An extensive software suite~\cite{SOFT-2022-02} is used in data simulation, in the reconstruction and analysis of real and simulated data, and in detector operations and the trigger and data acquisition systems of the experiment. All simulated Monte Carlo (MC) events are passed through a full simulation of the ATLAS detector response~\cite{SOFT-2010-01} using \GEANT~\cite{Agostinelli:2002hh}, and are corrected to match the number of interactions per bunch crossing (pile-up) observed in data. %
Only events satisfying requirements ensuring good detector operating conditions are analyzed.


%
\label{sec:MC}

Simulated VBF signal events were generated at next-to-leading order (NLO) in QCD with \POWHEGBOX[v2] \cite{Nason:2004rx,Frixione:2007vw,Alioli:2010xd,Nason:2009ai}, using the \NNPDF[3.0nlo]~\cite{Ball:2014uwa} parton distribution functions (PDFs). NLO electroweak (EW) corrections were calculated using \MGNLO~\cite{Alwall:2014hca} and applied as a function of the generated Higgs boson's transverse momentum (\pT). The gluon--gluon fusion (ggF) process can produce events with two energetic forward jets, resulting in a topology similar to that of VBF, and such events are treated as part of the signal. Simulated ggF events were generated at next-to-next-to-leading order (NNLO) in QCD with \POWHEGBOX[v2]~\cite{Hamilton:2013fea,Campbell:2012am,Hamilton:2015nsa} using the \NNPDF[3.0nlo] PDFs. Both the VBF and ggF samples used \PYTHIA[8]~\cite{Sjostrand:2014zea} with the A14 set of tuned parameters~\cite{ATL-PHYS-PUB-2014-021} and the \NNPDF[2.3lo] PDF set~\cite{Ball:2012cx} for parton showering and fragmentation. For the VBF sample, the \PYTHIA dipole-recoil scheme~\cite{dipoleRecoil} was used to improve the modeling of radiation in the central region. Other Higgs boson production modes, including both those with an associated vector boson ($VH$), top quark ($tH$) or top-quark pair (\ttH) and QCD-induced $b\bar{b}H$ production, are not considered due to their negligible contributions.

The dominant background, non-resonant QCD multijet production (Non-res.\ bkgd), is estimated using a data-driven method and accounts for about 99\% of the total background. Simulated multijet events are used only to derive the systematic uncertainty of this estimate. The multijet samples were generated with \PYTHIA[8] at leading order (LO) in QCD for inclusive $2 \to 2$ dijet production, using the \NNPDF[2.3lo] PDF set and the A14 tune. %
A filter was applied at generator level to enhance the fraction of events with a VBF-like topology. %

The hadronic resonant background from $V$+\:\!jets consists of QCD and EW components.
The QCD $V$+\:\!jets contribution was modeled at LO in QCD using \MGNLO with the \NNPDF[3.0nlo] PDF set and includes matrix elements with up to two partons at the Born level. The events were interfaced to \PYTHIA[8] for parton showering and hadronization.
The EW $V$+\:\!jets processes were modeled at LO in the EW coupling ($\mathcal{O}(\alpha_{\text{EW}}^4)$) by \SHERPA[2.2.14]~\cite{Bothmann:2019yzt} with the \NNPDF[3.0nlo] PDF set and include matrix elements with up to three partons at the Born level. The same filter as for the QCD multijet sample was applied at generator level to enhance the fraction of VBF-like events.
The EW $V$+\:\!jets component contributes most of the $V$+\:\!jets background in the most sensitive regions.
In the \Hbb\ regions, \Zbb\ dominates, while in the \Hcc\ regions, \Wcs\ and \Zcc\ are the main backgrounds, with the \Wcs\ to \Zcc\ ratio increasing from 0.40 in the most signal-like regions to 1.7 in the most background-like regions.
A small contribution from top-quark pair (\ttbar) production was modeled using \POWHEGBOX[v2] interfaced with \PYTHIA[8]. The \ttbar\ background contributes less than 10\% of the total $V$+\:\!jets yield in the most sensitive analysis regions. Except for \SHERPA\ samples, the decay of $b$ and $c$ hadrons was performed using \EVTGEN{1.6.0}~\cite{Lange:2001uf}.


%
\label{sec:eventReconstruction}


%
Jet reconstruction uses the particle-flow algorithm~\cite{PERF-2015-09}, which combines calorimeter energy deposits and inner-detector tracks into particle-flow objects that are clustered into jets using the anti-$k_t$ algorithm~\cite{Cacciari:2008gp,Fastjet} with a radius parameter $R=0.4$.
To mitigate the impact of \pileup, dedicated jet-vertex-tagging (JVT) requirements~\cite{ATLAS-CONF-2014-018} are applied. Jets with pseudorapidity $\abseta<2.5$ are required to have $\pT>20$~GeV, whereas jets with $\abseta>2.5$ must satisfy $\pT>30$~GeV.

Selected events must pass the inclusive VBF trigger~\cite{TRIG-2022-01, TRIG-2019-04}, which requires at least two jets: one with $\pT>70$~GeV and $\abseta<3.2$, and another with $\pT>50$~GeV and $\abseta<4.9$. Additional dijet requirements are imposed, namely $\detajj>4$, $\dphijj<2$, and $\mjj>1000~ (1100)$~GeV for \RunThr (\RunTwo). To avoid inefficiencies near the trigger threshold, the offline selection applies stricter requirements to the two VBF candidate jets (denoted by $j_1$ and $j_2$, ordered in decreasing jet \pT). Both jets must have \pT\ $>75$ GeV, with $j_1$ restricted to $\abseta<3.2$ and $j_2$ to $\abseta<4.5$. The dijet system is also required to satisfy $\detajonejtwo>4$, $\dphijonejtwo<2$, and $\mjonejtwo>1200$~GeV. When multiple jet pairs pass the requirements, the one with the largest \mjonejtwo\ is chosen.

The trigger efficiency is evaluated in data from a \Zmumu\ region, where events are first selected using a single-muon trigger~\cite{TRIG-2018-01}. This region is defined with selection criteria similar to those for the signal regions (SRs), except it requires two opposite-sign muons with $\pT> 7$~GeV and $\abseta< 2.5$, whose reconstruction and identification are described further below, instead of two $b$-tagged or $c$-tagged jets from Higgs boson decay. To emulate the kinematic topology of the signal region, events must also have $p_{\text{T},\ell\ell}> 150$~GeV and $|\Delta\phi_{\ell \ell}|< 2$. In these regions, the trigger efficiency is found to be above $90\%~ (95\%)$ in \RunTwo (\RunThr), with negligible dependence on jet kinematics.

Jets not identified as VBF candidate jets are considered for the selection of Higgs candidate jets (denoted by $j_{h1}$ and $j_{h2}$), originating from the \Hcc\ or \Hbb\ decay. Both $j_{h1}$ and $j_{h2}$ are required to have $\pT > 20$~GeV and $\abseta < 2.5$. To suppress shape distortions in the background invariant-mass distributions, the Higgs candidate jets must also satisfy $\pThonehtwo > 150$~GeV and $\dphihonehtwo < 2$. If multiple jets satisfy these criteria, the pair maximizing \pThonehtwo\ is selected.

The two Higgs candidate jets must be either $b$-tagged or $c$-tagged. Flavor identification is performed using the transformer-based GN2 algorithm~\cite{FTAG-2023-05}, which exploits charged-track information and achieves better separation of $b$-jets, $c$-jets, and light-flavor jets than the previous DL1r tagger~\cite{FTAG-2019-07}.
The calibrated tagger output defines working points corresponding to various $b$- and $c$-jet efficiency targets, using established calibration methods consistent with those described in Refs.~\cite{FTAG-2018-01, FTAG-2019-02, FTAG-2020-08}. To identify $b$-jets, this analysis uses the 77\% $b$-tagging-efficiency working point, which has a $c$-jet (light-flavor jet) efficiency of 6\% (0.24\%). Jets failing this requirement are classified as $c$-tagged if they satisfy the 50\% $c$-tagging-efficiency working point, which has a $b$-jet (light-flavor jet) efficiency of 13\% (5.3\%). Two tighter working points, corresponding to 30\% and 10\% $c$-jet efficiencies, provide additional inputs to the neural network. All quoted efficiency values are derived from simulated \ttbar\ events.

Events with one or more isolated electrons or muons are vetoed to avoid overlap with other \Hbb\ and \Hcc\ channels. Electrons are reconstructed by matching inner-detector tracks to electromagnetic-calorimeter energy clusters, while muons are reconstructed by combining inner-detector and muon-spectrometer tracks. Both electrons and muons must have $\pT>7$~GeV, $\abseta<2.5$, be matched to the primary vertex, and satisfy the loose identification and isolation criteria, as described in Refs.~\cite{EGAM-2021-01, MUON-2018-03}.

To recover energy lost in semileptonic heavy-flavor decays, the four-momentum of any muons with $\pT>3$~GeV that are reconstructed within the $b$- or $c$-tagged jet, prior to the muon isolation requirement, is added to the jet~\cite{HIGG-2020-20}. For $b$-jets, an additional $\pT$-dependent calibration is applied to account for residual response differences~\cite{HIGG-2020-20}.

Events are categorized into three channels: the $b\bar{b}$ Run-3 channel, where both Higgs candidate jets are $b$-tagged, and the $c\bar{c}$ Run-2 and Run-3 channels, where both are $c$-tagged. %


%
\label{sec:ANN for event classification}

%

%

The overall analysis strategy is similar to that in Ref.~\cite{HIGG-2019-04}. A neural-network classifier is used to separate events into different analysis regions, and the signal yield is extracted from a simultaneous fit to the Higgs candidate mass (\mhonehtwo) distribution in all regions. The dominant Non-res.\ bkgd is estimated from data; its shape is determined from the lowest-score region, assuming it is largely identical across all regions. To realize this strategy, an adversarial neural network (ANN), similar to that in Ref.~\cite{HIGG-2019-04}, is trained separately for each analysis channel in order to decorrelate the neural-network classifier output from \mhonehtwo. The networks are implemented in \textsc{PyTorch}~\cite{paszke2019pytorch} and optimized using the Adam algorithm~\cite{kingma2017adammethodstochasticoptimization}. Inputs comprise event and jet kinematics and jet flavor-tagging information. For training, signal events are taken from VBF MC samples, while background events are drawn from data sidebands on either side of the Higgs boson mass window ($100 < \mhonehtwo < 140~\GeV$) but within the range $50 < \mhonehtwo < 200~\GeV$. Further details of the ANN are provided in the End Matter. The number of regions and their boundaries are optimized for maximum sensitivity, resulting in six signal regions per channel (SR1--SR6), ordered by decreasing signal purity, which ranges from 2\% to 0.02\%. The least pure region, SR6, is used as a control region.

%

%

%


%
\label{sec:ZmumuCR}

%


%
\label{sec:Fit and systs}

The \Hcc\ contamination in the \Hbb\ channel is negligible, whereas the expected \Hbb\ yield in the \Hcc\ channel increases from about 40\% to 100\% of the \Hcc\ signal across SR1--SR5 and is approximately 1.4 times larger in SR6. The expected ggF-to-VBF yield ratio increases from about 1\% in SR1 to 8\% in SR5, and reaches about 45\% in SR6 in all channels; the ggF ANN distribution closely resembles that of the backgrounds, explaining the weak sensitivity of the measurement to ggF.

Signal production rates are extracted from a binned maximum-likelihood fit to the \mhonehtwo\ distributions in all regions and channels, using  5~\GeV bins over the range $50 <\mhonehtwo\ <200~\GeV$, resulting in a total of 30 bins per region. The expected \Hcc\ and \Hbb\ yields are scaled by independent signal-strength modifiers, $\mu^{c\bar{c}}$ and $\mu^{b\bar{b}}$, defined as the ratios of the observed to the SM-expected signal yields; being inclusive in production mode, they scale both the VBF and ggF contributions. These modifiers are treated as parameters of interest (POIs) and are extracted by maximizing the likelihood. Systematic uncertainties affecting both the normalization and shape of the fitted distributions are incorporated through nuisance parameters (NPs). Systematic variations with large statistical fluctuations are smoothed, while those with negligible impact are pruned region-by-region. %

The Non-res.\ bkgd is modeled using a data-driven approach. For each channel, its shape in the high-score regions (SR1--SR5) is taken from the signal-depleted low-score region (SR6). This strategy is motivated by the ANN design, which aims to decorrelate the reconstructed mass and classifier score. Technically, one unconstrained parameter is assigned per mass bin, correlated across all regions within a given channel, resulting in 90 such floating normalizations in total (one per bin per channel). Since SR6 contains a much larger number of events and is signal-depleted relative to SR1–SR5, it effectively constrains the multijet background shape in the high-score regions. In addition, 15 unconstrained normalization factors (one per high-score region per channel) are introduced, allowing the fit to freely adjust the QCD background yield in each region.

Two sets of systematic uncertainties are introduced to account for possible Non-res.\ bkgd shape differences between signal regions, arising from an imperfect decorrelation of the ANN score and \mhonehtwo. The \textit{residual uncertainties} capture potential correlations: for each signal region, an uncertainty is derived from \mhonehtwo\ sidebands and applied across the full mass range. A fit to the ratio of the \mhonehtwo\ distributions in the sidebands of SR6 to those in each of SR1--SR5 is performed, where a linear function adequately describes the observed differences; higher-order polynomials were tested but did not provide any significant improvement. The fitted slopes are implemented as unconstrained shape variations in the likelihood, while the offsets are absorbed by the normalization factors.

The \textit{bias uncertainties} quantify possible distortions of the \mhonehtwo\ spectrum within the Higgs mass window ($100 < \mhonehtwo < 140~\GeV$) that may arise because the ANN is trained only on sideband events and may bias the signal extraction. They are estimated by performing signal-plus-background fits on multijet MC samples, which are treated as data after being reweighted using sideband data information to improve the modelling. The resulting bias in the fitted signal strength, computed as the quadrature sum of the extracted signal yield and its statistical uncertainty, is taken as a systematic uncertainty. The bias amounts to about 27\%--51\% of the expected statistical uncertainty in the \Hbb\ signal and 26\%--56\% of that in the \Hcc\ signal.
Both the \textit{bias} and \textit{residual} uncertainties are applied in SR1--SR5, treated as uncorrelated across regions, leading to 15 nuisance parameters for each set.

Normalization factors for the $V$+\;\!jets backgrounds are extracted from the Run-2 and Run-3 \Zmumu\ regions. The \Zmumu\;\!+\;\!jets MC samples are simulated using the same setup as for the hadronic $V$+\:\!jets samples. The ANNs trained in the \Hcc\ channels are applied to these regions, replacing the $c$-jet candidates with the muon pair (with a fixed value of the $c$-tagging score to mimic $c$-jets), to define 12 score regions (6 in each Run). A dedicated statistical-only, normalization-only fit is performed simultaneously across all these regions to extract normalization factors for the QCD and EW $V$+\;\!jets components, corresponding to the background- and signal-like contributions, respectively. The resulting normalization factors are applied to simulated $V$+\;\!jets templates in the signal regions. A $50\%$ normalization uncertainty is applied to the sum of EW and QCD components, treated as uncorrelated between the $Z$\;\!+\;\!jets and $W$+\;\!jets backgrounds. The size of this uncertainty corresponds to the largest data–MC discrepancy observed in the \Zmumu\ region after the fit. A $50\%$ normalization uncertainty is also assigned to the \ttbar\ background. These uncertainties are treated as uncorrelated across regions.

An overall 7\% (4\%) uncertainty in the VBF trigger selection efficiency, derived from a comparison between data and simulation in the \Zmumu\ region, is applied to the Run-2 (Run-3) simulated events. Systematic uncertainties considered for the jets include those for the energy scale and resolution~\cite{JETM-2018-05}, the vertex-tagging efficiency~\cite{ATLAS-CONF-2014-018, ATL-PHYS-PUB-2019-026}, and the $b$/$c$-tagging efficiency for $b$-jets~\cite{FTAG-2018-01}, $c$-jets~\cite{FTAG-2020-08}, and light-flavor jets~\cite{FTAG-2019-02}. Uncertainties of 0.83\% and 2\% are applied to the integrated luminosity of the Run-2 and Run-3 datasets, respectively~\cite{DAPR-2021-01}.

For both VBF and ggF signal, uncertainties in the predicted SM production cross-sections and branching ratios follow the recommendations of Ref.~\cite{deFlorian:2016spz}. PDF uncertainties are evaluated by using eigenvector variations of the \NNPDF[2.3lo] set~\cite{Ball:2012cx}. The impact of the ggF process modeling in the VBF phase space was evaluated and found to be negligible. Consequently, no associated uncertainties are considered. In addition, for the VBF signal, QCD scale uncertainties are estimated by varying the renormalization and factorization scales, while underlying-event and parton-shower uncertainties are assessed by comparing simulations from \PYTHIA[8] and \HERWIG[7]~\cite{Sjostrand:2014zea, Bahr:2008pv}. The uncertainty from the EW NLO reweighting of the VBF signal was evaluated and found to have a negligible impact on the results.


%
\label{sec:result}

\begin{figure}
\begin{center}
\subfloat[]{\includegraphics[width=0.45\textwidth]{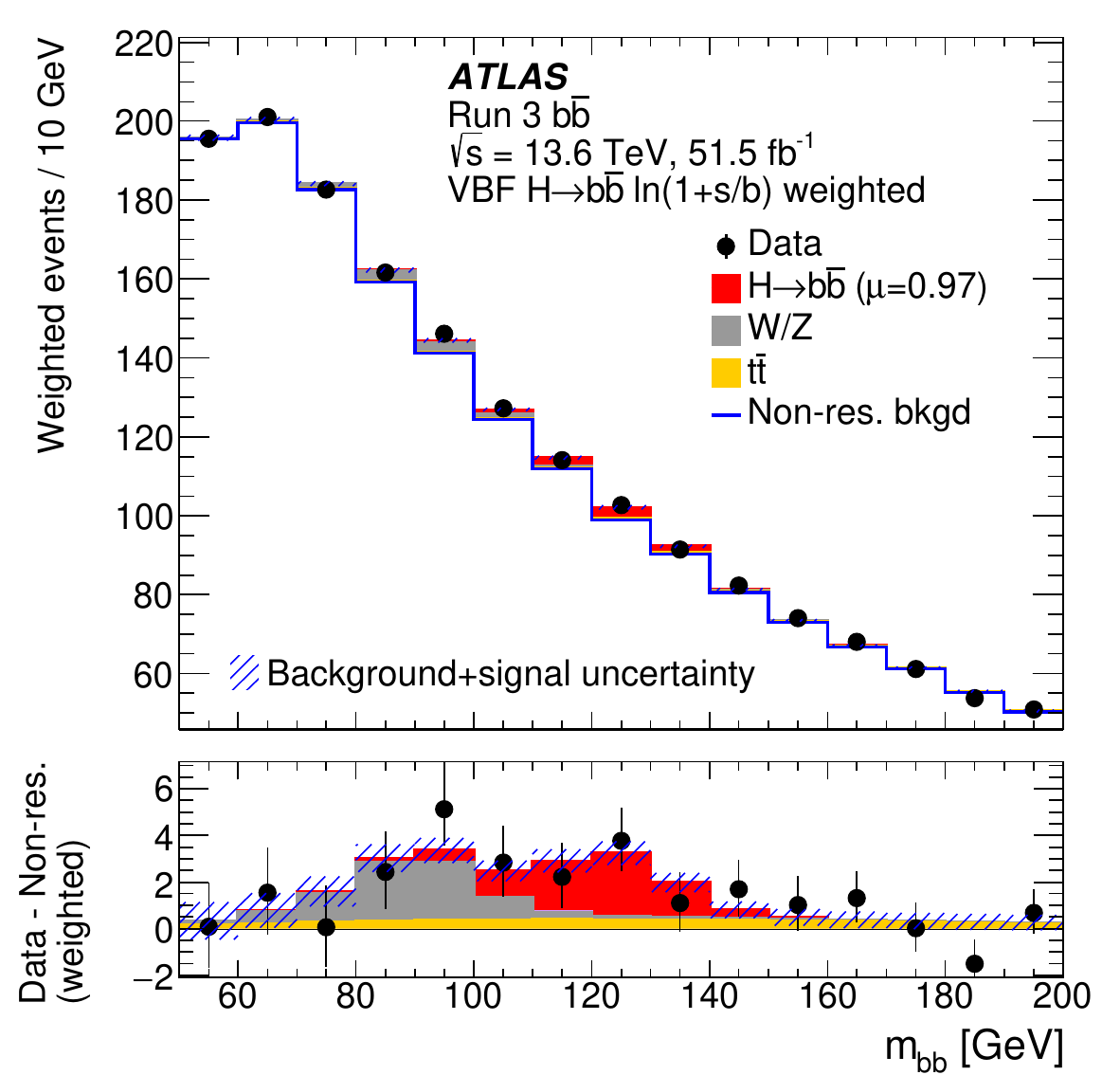}}
\subfloat[]{\includegraphics[width=0.45\textwidth]{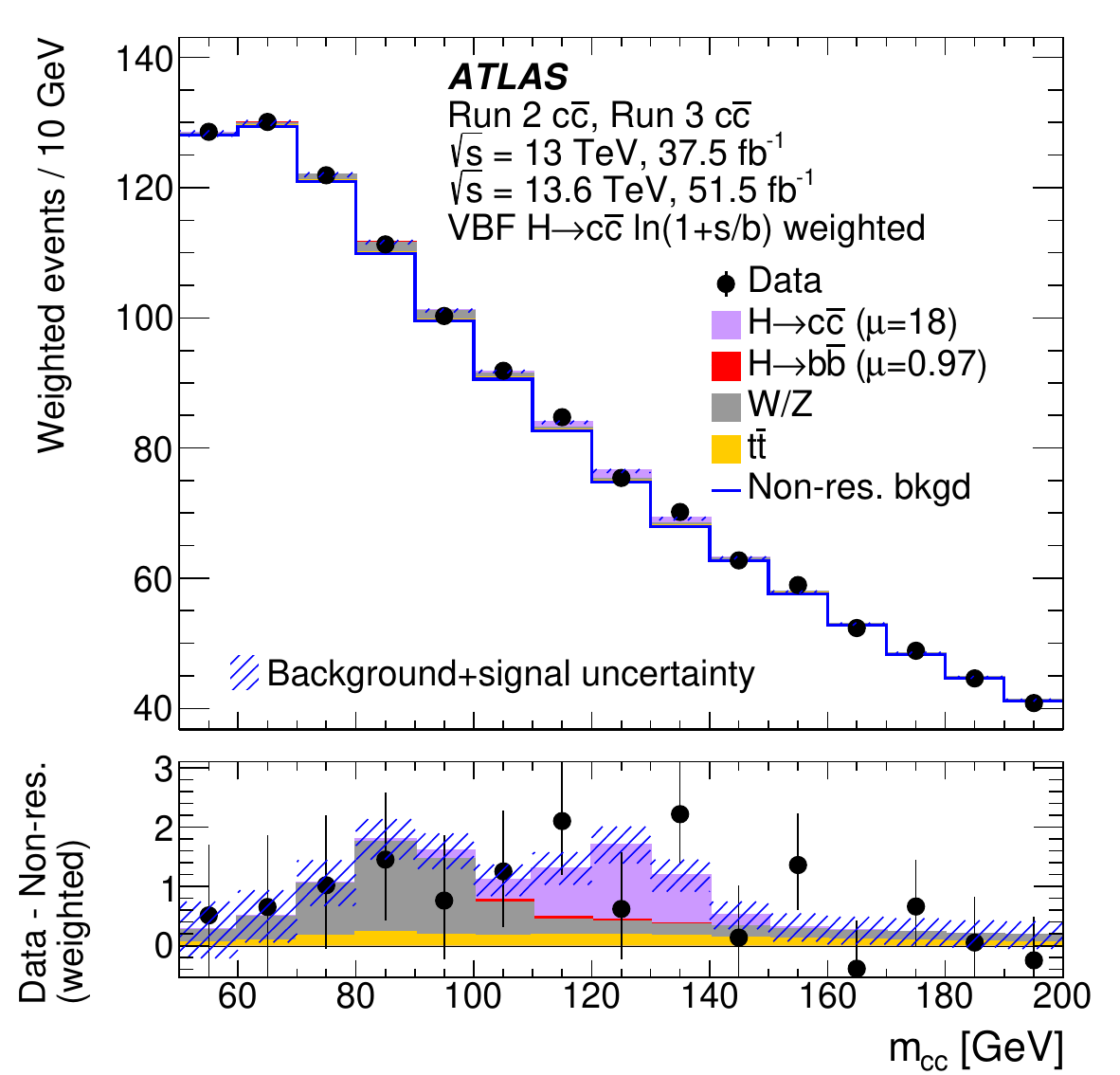}}
\caption{Higgs candidate mass distribution with all SRs weighted by $\ln(1+s/b)$ and summed for the (a) \Hbb\ channel and (b) \Hcc\ channel. The value of $s$ ($b$) is calculated from the observed Higgs boson signal (post-fit background yields) in each region. The bottom panel shows the difference between the weighted data and the Non-res.\ bkgd. The diagonally striped band represents the total uncertainty of the fitted background and signal.}
\label{fig:databkg}
\end{center}
\end{figure}

\Fig{\ref{fig:databkg}} shows the observed data together with the signal and background estimates in each bin of the Higgs candidate mass discriminant. All SRs are weighted by $\ln(1+s/b)$ and summed; here $b$ and $s$ denote the total background and signal yields, respectively. The distributions are shown separately for the \Hbb\ and \Hcc\ channels, with the signal definition restricted to one Higgs boson decay mode at a time.

The measured signal strengths for the \Hbb\ and \Hcc\ signals are:
\begin{align*}
\mu^{b\bar{b}} &= \muVBFHbb ^{+0.57}_{-0.50} = \muVBFHbb \pm 0.45\:(\textrm{stat.}) ^{+0.35}_{-0.22}\:(\textrm{syst.})\\
\mu^{c\bar{c}} &= \muVBFHcc \pm 13 = \muVBFHcc \pm 12 \:(\textrm{stat.}) \pm 7 \:(\textrm{syst.})
\end{align*}
with a correlation of $-4.5\%$ between the two measurements. The measurement of $\mu^{b\bar{b}}$ yields an observed (expected) significance of 2.0 (2.1) standard deviations over the background-only prediction. No significant excess over the background-only hypothesis is observed in the search for \Hcc. An upper limit on $\mu^{c\bar{c}}$ is calculated using the CL$_\text{s}$ method~\cite{Read:2002hq}, with the profile-likelihood ratio as the test statistic~\cite{Cowan:2010js}. The observed 95\% confidence level (CL) upper limit on $\mu^{c\bar{c}}$ is \VBFHccLimObs, while a limit of \VBFHccLimExp\ is expected in the case of no \Hcc\ process. The result is also interpreted in the \enquote{kappa framework}~\cite{Heinemeyer:2013tqa,deFlorian:2016spz} by reparameterizing $\mu^{c\bar{c}}$ in terms of the charm Yukawa coupling modifier $\kappa_c$, defined as $\kappa_c^2 = \Gamma_c/\Gamma_c^{\text{SM}}$, where $\Gamma_c$ is the partial decay width into a charm-quark pair. Assuming that all other couplings follow their SM predictions and only SM decays are considered, the parameterization is:
\begin{eqnarray*}
\mu^{c\bar{c}} = \frac{\kappa_c^2}{1+B^{\text{SM}}_{\Hcc}(\kappa_c^2-1)}  \label{eq:kc}
\end{eqnarray*}
where $B^{\text{SM}}_{\Hcc}$ is the SM \Hcc\ branching fraction. The expected 95\% CL constraint on the modifier is $|\kappa_c| < 11.6$. No observed constraint can be set at 95\% CL due to the large observed value of $\mu^{c\bar{c}}$. Instead, a 68\% CL interval of $2.3 < |\kappa_c| < 18.4$ is obtained. An additional fit, performed by considering only the VBF process as signal, while fixing the ggF contribution to its SM prediction within uncertainties, yields a nearly identical result for both the \Hbb\ and \Hcc\ measurements. %

The contributions of each uncertainty source to the total uncertainty in the fitted $\mu^{c\bar{c}}$ and $\mu^{b\bar{b}}$ values are summarized in~\Tab{\ref{tab:break}}. Both measurements are statistically limited. In both cases, the dominant systematic uncertainties arise from the Non-res.\ bkgd bias uncertainty and the effects of limited MC sample size (particularly those from the QCD $V$+\;\!jets samples), followed by the detector-related uncertainties, primarily those associated with jets.

\begin{table}[hbtp]
\caption{Breakdown of contributions to the uncertainty in the fitted signal strengths for \Hcc\ and \Hbb. All values correspond to absolute uncertainties in $\mu^{c\bar{c}}$ and $\mu^{b\bar{b}}$. The sum in quadrature of uncertainties from different sources may differ from the total because of correlations. The mean of the absolute values of the upward and downward systematic variations is shown. \enquote{Other Higgs signal normalization} indicates an effect of $\mu^{b\bar{b}}$ ($\mu^{c\bar{c}}$) on the $\mu^{c\bar{c}}$ ($\mu^{b\bar{b}}$) measurement.}
\centering
\begin{tabular}{ l c c}
\toprule
Source of uncertainty & $\sigma(\mu^{c\bar{c}})$ & $\sigma(\mu^{b\bar{b}})$ \\
Total                 & 13~~      & 0.53        \\
\hline
Statistical           & 12~~      & 0.45        \\
~~~Data statistical  & 8.7       & 0.32        \\
~~~Non-res.\ bkgd normalization & 4.9       & 0.19        \\
~~~Non-res.\ bkgd shape & 4.6       & 0.20        \\
~~~Non-res.\ bkgd residual uncertainty & 1.4       &   0.05      \\
~~~Other Higgs signal normalization &  0.3      & 0.01      \\
\hline
Systematic            & 6.5      & 0.28        \\
~~~Non-res.\ bkgd bias uncertainty & 5.8       & 0.19       \\
~~~Simulation sample sizes & 3.4       & 0.21       \\
~~~Experimental &  1.8        & 0.18       \\
~~~~~Jet &  1.8        & 0.18      \\
~~~~~Flavor tagging &  0.3        & 0.01        \\
~~~~~Trigger &  0.3       & 0.02       \\
~~~~~Other & ~~0.06        & 0.02       \\
~~~$V$+\;\!jets and \ttbar\ background estimate & 0.7       & 0.03        \\
~~~Signal theory &   0.8      & 0.04       \\
\bottomrule
\end{tabular}
\label{tab:break}
\end{table}

Two combined analyses are performed: one combining the \Hcc\ result with the Run-2 $VH$-channel results~\cite{HIGG-2020-20}, and another combining the Run-3 \Hbb\ result with the Run-2 result in the VBF channel~\cite{HIGG-2019-04} and a recently reported Run-2 photon-associated VBF (VBF\;\!+\;\!$\gamma$) result~\cite{HIGP-2024-04}. For the \Hcc\ combination, $\mu^{b\bar{b}}$ is also allowed to float. For the \Hbb\ combination, the \Hcc\ channels from this analysis are excluded and $\mu^{c\bar{c}}$ is fixed to 1, with the ggF contributions fixed to their SM predictions within uncertainties. For both combinations, jet-related experimental uncertainties are correlated, except for those related to jet flavor tagging, where different algorithms and calibration procedures are used; background modeling uncertainties are treated as uncorrelated, while theory uncertainties in signal production cross-sections and branching ratios are correlated.

From the first combination, the observed $\mu_\text{Comb.}^{c\bar{c}}$ signal strength relative to the SM prediction is $3.3^{+5.2}_{-5.0}$, assuming SM VBF, ggF, and $VH$ production rates. The observed (expected) 95\% CL upper limit on $\mu_\text{Comb.}^{c\bar{c}}$ is \hccLimObs\ (\hccLimExp), corresponding to an observed (expected) constraint of $|\kappa_c| < \kappacObs ~(\kappacExp)$. This represents an 8\% improvement in the expected limit and a 5\% improvement in the expected $|\kappa_c|$ constraint relative to the $VH$-only result. The observed result is slightly weaker than that from the $VH$ channel, owing to an upward fluctuation in $\mu_\text{Comb.}^{c\bar{c}}$ driven by the VBF contribution. From the second combination, the observed $\mu_\text{VBF}^{b\bar{b}}$ signal strength relative to the SM prediction is $0.87 \pm 0.28$, corresponding to an observed (expected) significance of 3.2 (3.6) standard deviations, thus providing the first evidence for the \vbfhbb process. The $p$-value for compatibility of the signal-strength parameters among the three analyses is 65\%, and 95\% between the Run-2 and Run-3 \vbfhbb\ analyses.

\FloatBarrier


%

%
%
%

%
%
\label{sec:conclusion}

%
%

In summary, this Letter presents the first search for SM Higgs boson decay into a pair of charm quarks in the VBF production mode in the ATLAS experiment at the LHC, together with a simultaneous measurement of Higgs boson decay into a bottom quark--antiquark pair. The \vbfhcc\ search is enabled by a new trigger designed to exploit the VBF topology. The observed signal strengths relative to the SM predictions are $\mu^{b\bar{b}} = \muVBFHbb^{+0.57}_{-0.50}$ and $\mu^{c\bar{c}} = \muVBFHcc \pm 13$. For the \Hcc\ process, this corresponds to an observed (expected) 95\% CL upper limit of \VBFHccLimObs\ (\VBFHccLimExp) on $\mu^{c\bar{c}}$. When combined with previous Run-2 results, these measurements improve the expected constraint on the Higgs–charm coupling and both the expected and observed constraints on the Higgs–bottom coupling.%


%

%
%
%
%
%
\section*{Acknowledgments}
%

%

%
%

%
%

We thank CERN for the very successful operation of the LHC and its injectors, as well as the support staff at
CERN and at our institutions worldwide without whom ATLAS could not be operated efficiently.

The crucial computing support from all WLCG partners is acknowledged gratefully, in particular from CERN, the ATLAS Tier-1 facilities at TRIUMF/SFU (Canada), NDGF (Denmark, Norway, Sweden), CC-IN2P3 (France), KIT/GridKA (Germany), INFN-CNAF (Italy), NL-T1 (Netherlands), PIC (Spain), RAL (UK) and BNL (USA), the Tier-2 facilities worldwide and large non-WLCG resource providers. Major contributors of computing resources are listed in Ref.~\cite{ATL-SOFT-PUB-2025-001}.

We gratefully acknowledge the support of ANPCyT, Argentina; YerPhI, Armenia; ARC, Australia; BMWFW and FWF, Austria; ANAS, Azerbaijan; CNPq and FAPESP, Brazil; NSERC, NRC and CFI, Canada; CERN; ANID, Chile; CAS, MOST and NSFC, China; Minciencias, Colombia; MEYS CR, Czech Republic; DNRF and DNSRC, Denmark; IN2P3-CNRS and CEA-DRF/IRFU, France; SRNSFG, Georgia; BMFTR, HGF and MPG, Germany; GSRI, Greece; RGC and Hong Kong SAR, China; ICHEP and Academy of Sciences and Humanities, Israel; INFN, Italy; MEXT and JSPS, Japan; CNRST, Morocco; NWO, Netherlands; RCN, Norway; MNiSW, Poland; FCT, Portugal; MNE/IFA, Romania; MSTDI, Serbia; MSSR, Slovakia; ARIS and MVZI, Slovenia; DSI/NRF, South Africa; MICIU/AEI, Spain; SRC and Wallenberg Foundation, Sweden; SERI, SNSF and Cantons of Bern and Geneva, Switzerland; NSTC, Taipei; TENMAK, T\"urkiye; STFC/UKRI, United Kingdom; DOE and NSF, United States of America.

Individual groups and members have received support from BCKDF, CANARIE, CRC and DRAC, Canada; CERN-CZ, FORTE and PRIMUS, Czech Republic; COST, ERC, ERDF, Horizon 2020 and Marie Sk{\l}odowska-Curie Actions, European Union; Investissements d'Avenir Labex, Investissements d'Avenir Idex and ANR, France; DFG and AvH Foundation, Germany; Herakleitos, Thales and Aristeia programmes co-financed by EU-ESF and the Greek NSRF, Greece; BSF-NSF and MINERVA, Israel; NCN and NAWA, Poland; La Caixa Banking Foundation, CERCA Programme Generalitat de Catalunya and PROMETEO and GenT Programmes Generalitat Valenciana, Spain; G\"{o}ran Gustafssons Stiftelse, Sweden; The Royal Society and Leverhulme Trust, United Kingdom.

In addition, individual members wish to acknowledge support from Chile: Agencia Nacional de Investigaci\'on y Desarrollo (FONDECYT 1230812, FONDECYT 1240864, Fondecyt 3240661, Fondecyt Regular 1240721); China: Chinese Ministry of Science and Technology (MOST-2023YFA1605700, MOST-2023YFA1609300), National Natural Science Foundation of China (NSFC - 12175119, NSFC 12275265); Czech Republic: Czech Science Foundation (GACR - 24-11373S), Ministry of Education Youth and Sports (ERC-CZ-LL2327, FORTE CZ.02.01.01/00/22\_008/0004632), PRIMUS Research Programme (PRIMUS/21/SCI/017); EU: H2020 European Research Council (ERC - 101002463); European Union: European Research Council (BARD No. 101116429, ERC - 948254, ERC 101089007), European Regional Development Fund (SMASH COFUND 101081355, SLO ERDF), European Union, Future Artificial Intelligence Research (FAIR-NextGenerationEU PE00000013); France: Agence Nationale de la Recherche (ANR-21-CE31-0022, ANR-22-EDIR-0002, ANR-24-CE31-0504-01); Germany: Deutsche Forschungsgemeinschaft (DFG - 469666862, DFG - CR 312/5-2); China: Research Grants Council (GRF); Italy: Ministero dell'Università e della Ricerca (NextGenEU 153D23001490006 M4C2.1.1, NextGenEU I53D23000820006 M4C2.1.1, NextGenEU I53D23001490006 M4C2.1.1, SOE2024\_0000023); Japan: Japan Society for the Promotion of Science (JSPS KAKENHI JP22H01227, JSPS KAKENHI JP22H04944, JSPS KAKENHI JP22KK0227, JSPS KAKENHI JP24K23939, JSPS KAKENHI JP24KK0251, JSPS KAKENHI JP25H00650, JSPS KAKENHI JP25H01291, JSPS KAKENHI JP25K01023); Norway: Research Council of Norway (RCN-314472); Poland: Ministry of Science and Higher Education (IDUB AGH, POB8, D4 no 9722), Polish National Science Centre (NCN 2021/42/E/ST2/00350, NCN OPUS 2023/51/B/ST2/02507, NCN UMO-2019/34/E/ST2/00393, UMO-2022/47/O/ST2/00148, UMO-2023/49/B/ST2/04085, UMO-2023/51/B/ST2/00920, UMO-2024/53/N/ST2/00869); Portugal: Foundation for Science and Technology (FCT); Spain: Generalitat Valenciana (ASFAE/2022/008), Ministry of Science and Innovation (RYC2019-028510-I, RYC2020-030254-I, RYC2021-031273-I, RYC2022-038164-I), Ministerio de Ciencia, Innovación y Universidades/Agencia Estatal de Investigaci\'on (PID2022-142604OB-C22); Sweden: Carl Trygger Foundation (Carl Trygger Foundation CTS 22:2312), Swedish Research Council (Swedish Research Council 2023-04654, VR 2021-03651, VR 2022-03845, VR 2022-04683, VR 2023-03403, VR 2024-05451), Knut and Alice Wallenberg Foundation (KAW 2018.0458, KAW 2022.0358, KAW 2023.0366); Switzerland: Swiss National Science Foundation (SNSF - PCEFP2\_194658); United Kingdom: The Binks Trust, Royal Society (NIF-R1-231091); United States of America: U.S. Department of Energy (ECA DE-AC02-76SF00515), John Templeton Foundation (John Templeton Foundation 63206), Neubauer Family Foundation.

%
%


%
\clearpage
\appendix
\part*{End Matter}
\addcontentsline{toc}{part}{End Matter}
\label{sec:appendix}

\begin{figure}[htbp]
\begin{center}
\subfloat[]{\includegraphics[width=0.3\textwidth]{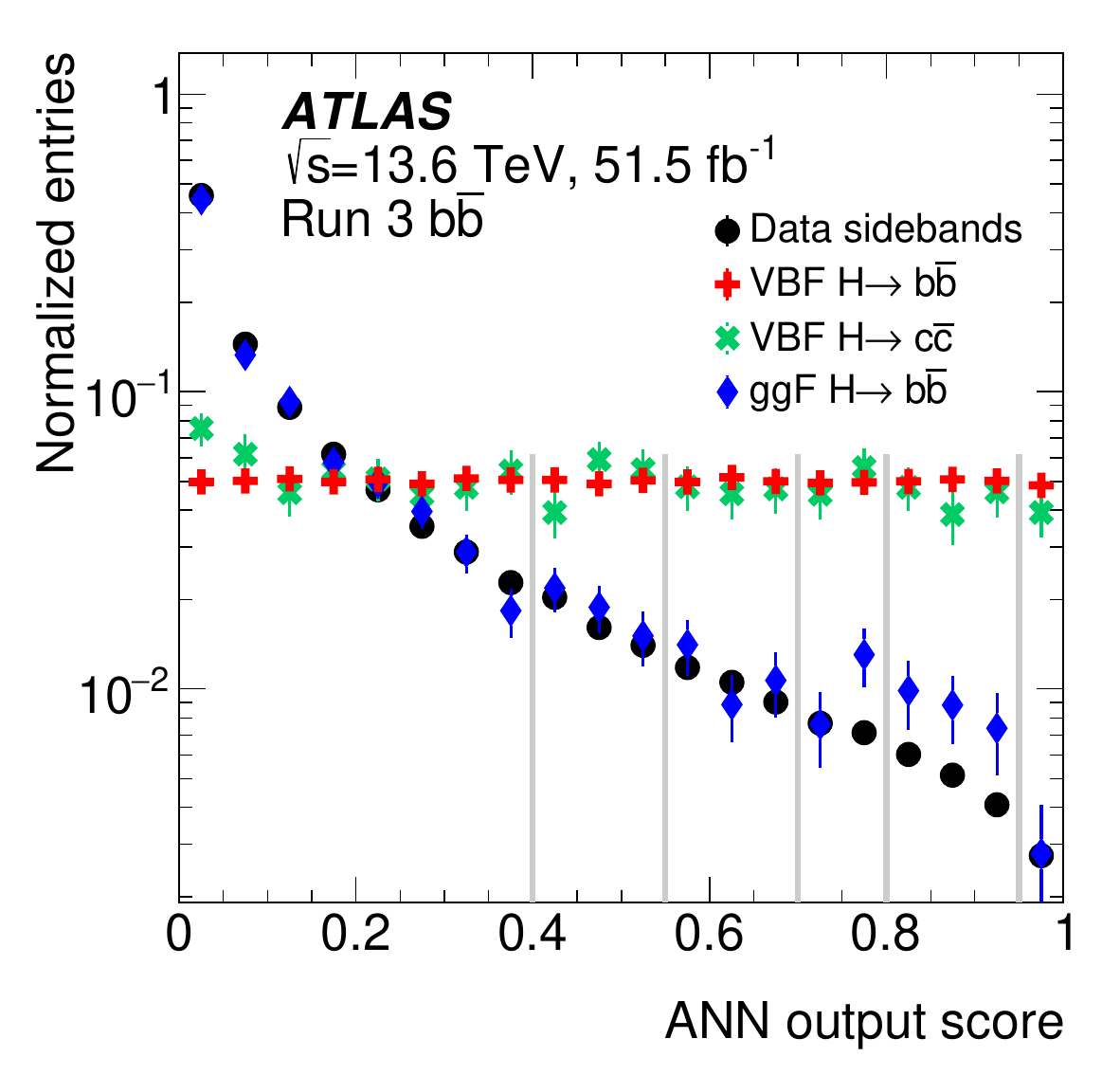}}
\subfloat[]{\includegraphics[width=0.3\textwidth]{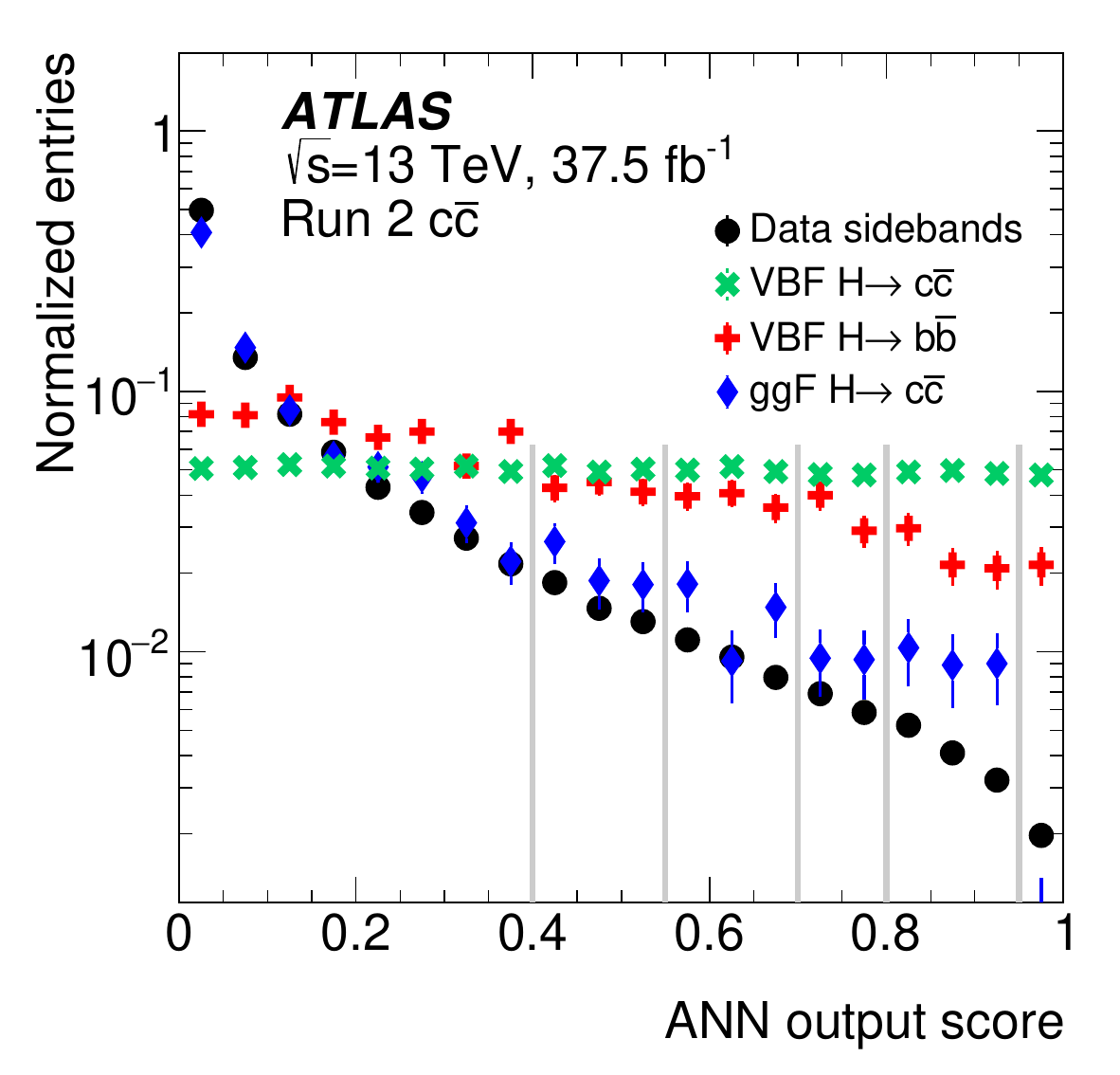}}
\subfloat[]{\includegraphics[width=0.3\textwidth]{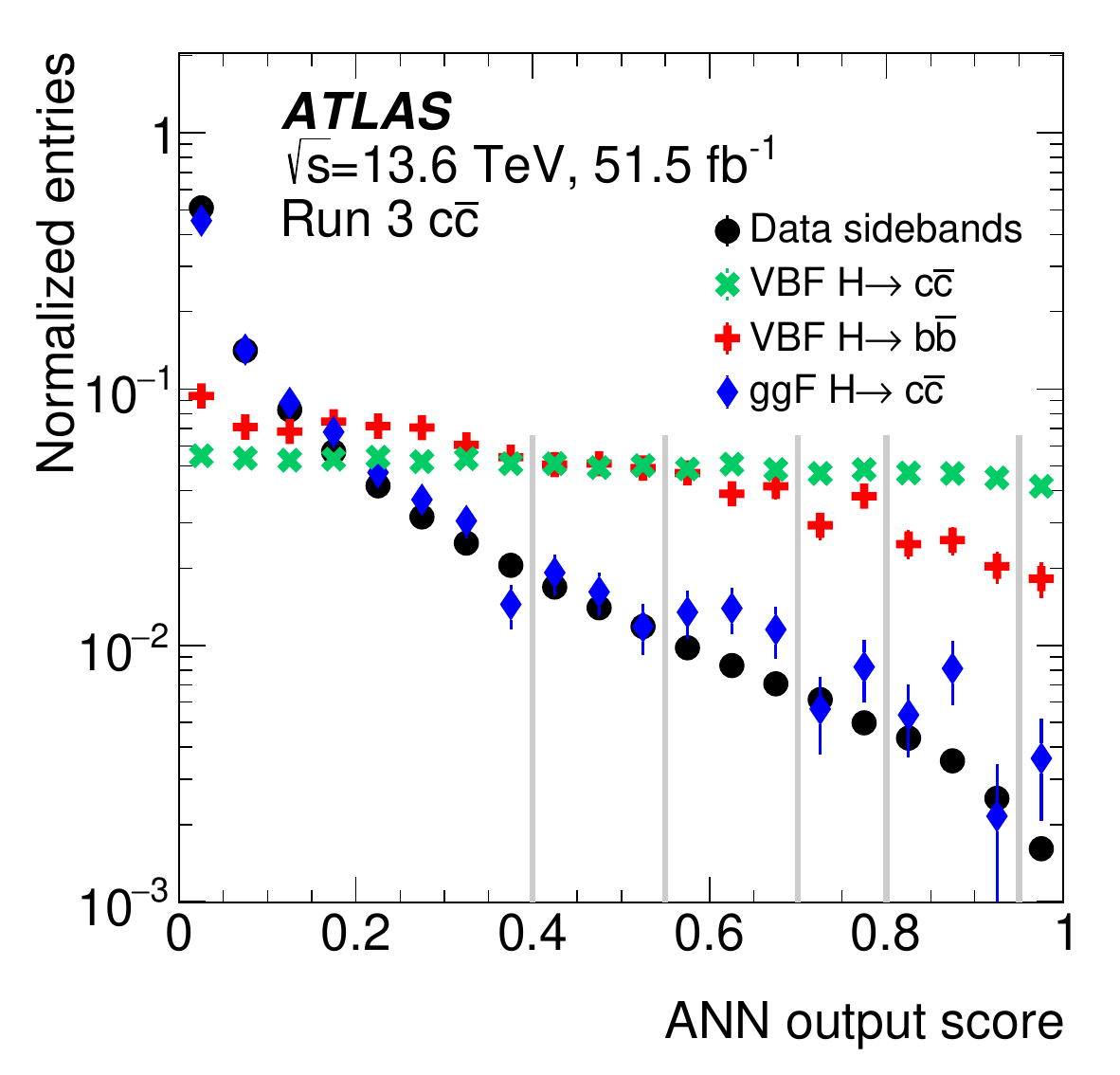}}\\
\caption{Distributions of ANN output scores for the \vbfhcc\ signal MC events, \vbfhbb\ signal MC events, ggF MC events, and data sidebands in the (a) $b\bar{b}$ Run-3, (b) $c\bar{c}$ Run-2, and (c) $c\bar{c}$ Run-3 channels. The errors are statistical only. Vertical lines indicate the ANN score thresholds defining the six signal regions (SR1--SR6), in right-to-left order, with the rightmost region SR1 being the most signal-like. No separation between $b\bar{b}$ and $c\bar{c}$ signals is expected in the $b\bar{b}$ ANN since tagging information is not used in this channel, while such separation is visible in the $c\bar{c}$ ANN, which includes tagging inputs.}
\label{fig:ANN}
\end{center}
\end{figure}

The input variables used in ANN training are listed in Table~\ref{tab:Inputs}. The \Hbb\ and \Hcc\ channels use almost identical sets of variables, differing only by the inclusion of tagging-related inputs in the \Hcc\ channels, which improve discrimination between $c$-jets and background, and the \pTadd\ variable, defined as the transverse momentum of the leading additional jet, in the \Hbb\ channel. The $c\bar{c}$ Run-2 and Run-3 channels share an identical input set. Figure~\ref{fig:ANN} shows the ANN score distributions for the VBF signal, ggF events, and the data sidebands. As a post-training step, the classifier scores are scaled to quantiles of the signal distribution with the output values ranging from 0 to 1. Clear separation is visible between the VBF signal and the data sidebands, while the ggF events exhibit a score distribution similar to that of the data sidebands. %
The variables that contribute most significantly to the ANN's discriminating power are \mjonejtwo, \pTjonejtwo, \pTbal and \nJetsRap.

\begin{table}[hbtp]
\caption{The input variables used for ANN training in each of the channels. The \checkmark\ symbol indicates the inclusion of a variable.}
\label{tab:Inputs}
\centering
\begin{tabularx}{\textwidth}{lccX}
\toprule
& \Hcc\ & \Hbb & \\
\hline
Variable & & & Description \\
\hline\hline
\mjonejtwo&  \checkmark\ & \checkmark\ & Invariant mass of the VBF candidate jet pair.\\
\hline
\pTjonejtwo& \checkmark\ & \checkmark\ & Transverse momentum of the VBF candidate jet pair.\\
\hline
\pThonehtwo& \checkmark\ & \checkmark\ & Transverse momentum of the Higgs jet pair.\\
\hline
&&&\\[-2.4ex]
\pTbal     & \checkmark\ & \checkmark\ & The ratio of the vectorial to scalar sums of the transverse momenta of $j_{h1},j_{h2}, j_1$ and $j_2$.\\
\hline
&&&\\[-2.4ex]
$\theta_{j_{h}j_{h}}$ & \checkmark\ & \checkmark\ &  Defined as $\tan^{-1} \left[ \tan \left(\frac{\Delta\eta(j_{h}j_{h})}{2}\right) / \tanh \left(\frac{\Delta\phi(j_{h}j_{h})}{2}\right) \right]$, a measure of the relative angle of $\Delta\eta$ and $\Delta\phi$ between the two tagged jets. \\
\hline
&&&\\[-2.4ex]
\nJetsRap &  \checkmark\ & \checkmark\ & The number of jets with $\pT > 20\,\GeV$ and $\eta$ between $\eta_{j_1}$ and $\eta_{j_2}$. \\
\hline
&&&\\[-2.4ex]
\dPhihhjj &  \checkmark\ & \checkmark\ & The separation in $\phi$ between the Higgs jet pair and the VBF candidate jet pair.\\
\hline
&&&\\[-2.4ex]
$N_{\text{trk}}^{j_{1}}$ &  \checkmark\ & \checkmark\ & The number of tracks matched to the leading VBF candidate jet. \\
\hline
&&&\\[-2.4ex]
$N_{\text{trk}}^{j_{2}}$ &  \checkmark\ & \checkmark\ & The number of tracks matched to the sub-leading VBF candidate jet. \\
\hline
&&&\\[-2.4ex]
$D_{\text{GN2}}^{j_{h1}}$ &  \checkmark\ &  & Binned output of the GN2 tagging algorithm for the leading $c$-tagged jet, with bin boundaries corresponding to $50\%$, $30\%$, and $10\%$ $c$-tagging efficiencies. \\
\hline
&&&\\[-2.4ex]
$D_{\text{GN2}}^{j_{h2}}$ &  \checkmark\ &  & Binned output of the GN2 tagging algorithm for the sub-leading $c$-tagged jet, with bin boundaries corresponding to $50\%$, $30\%$, and $10\%$ $c$-tagging efficiencies. \\
\hline
&&&\\[-2.4ex]
\pTadd & & \checkmark\ & Transverse momentum of the additional jet with the highest transverse momentum that does not belong to either the VBF candidate jets or the Higgs jets, and satisfies $|\eta| < 2.5$. If no such jet exists, the value is set to 0 by default. \\
\bottomrule
\end{tabularx}
\end{table}

\FloatBarrier


%
%

%
%
%
\printbibliography
\clearpage
\input{atlas_authlist}
%

%
\clearpage

\end{document}

%% file: atlas_authlist.tex
 
\begin{flushleft}
\hypersetup{urlcolor=black}
{\Large The ATLAS Collaboration}

\bigskip

\AtlasOrcid[0000-0002-6665-4934]{G.~Aad}$^\textrm{\scriptsize 103}$,
\AtlasOrcid[0000-0001-7616-1554]{E.~Aakvaag}$^\textrm{\scriptsize 17}$,
\AtlasOrcid[0000-0002-5888-2734]{B.~Abbott}$^\textrm{\scriptsize 122}$,
\AtlasOrcid[0000-0002-0287-5869]{S.~Abdelhameed}$^\textrm{\scriptsize 118a}$,
\AtlasOrcid[0000-0002-1002-1652]{K.~Abeling}$^\textrm{\scriptsize 55}$,
\AtlasOrcid[0000-0001-5763-2760]{N.J.~Abicht}$^\textrm{\scriptsize 49}$,
\AtlasOrcid[0000-0002-8496-9294]{S.H.~Abidi}$^\textrm{\scriptsize 30}$,
\AtlasOrcid[0009-0003-6578-220X]{M.~Aboelela}$^\textrm{\scriptsize 45}$,
\AtlasOrcid[0000-0002-9987-2292]{A.~Aboulhorma}$^\textrm{\scriptsize 36e}$,
\AtlasOrcid[0000-0001-5329-6640]{H.~Abramowicz}$^\textrm{\scriptsize 156}$
% ...
\bigskip
\\

% ...
$^{17}$Department for Physics and Technology, University of Bergen, Bergen; Norway.\\
% ...
$^{30}$Physics Department, Brookhaven National Laboratory, Upton NY; United States of America.\\
% ...
$^{36}$$^{(a)}$Facult\'e des Sciences Ain Chock, Universit\'e Hassan II de Casablanca;$^{(b)}$Facult\'{e} des Sciences, Universit\'{e} Ibn-Tofail, K\'{e}nitra;$^{(c)}$Facult\'e des Sciences Semlalia, Universit\'e Cadi Ayyad, LPHEA-Marrakech;$^{(d)}$LPMR, Facult\'e des Sciences, Universit\'e Mohamed Premier, Oujda;$^{(e)}$Facult\'e des sciences, Universit\'e Mohammed V, Rabat;$^{(f)}$Institute of Applied Physics, Mohammed VI Polytechnic University, Ben Guerir; Morocco.\\
% ...
$^{45}$Physics Department, Southern Methodist University, Dallas TX; United States of America.\\
% ...
$^{49}$Fakult\"{a}t Physik, Technische Universit{\"a}t Dortmund, Dortmund; Germany.\\
% ...
$^{55}$II. Physikalisches Institut, Georg-August-Universit\"{a}t G\"ottingen, G\"ottingen; Germany.\\
% ...
$^{103}$CPPM, Aix-Marseille Universit\'e, CNRS/IN2P3, Marseille; France.\\
% ...
$^{118}$$^{(a)}$New York University Abu Dhabi, Abu Dhabi;$^{(b)}$United Arab Emirates University, Al Ain; United Arab Emirates.\\
% ...
$^{122}$Homer L. Dodge Department of Physics and Astronomy, University of Oklahoma, Norman OK; United States of America.\\
% ...
$^{156}$Raymond and Beverly Sackler School of Physics and Astronomy, Tel Aviv University, Tel Aviv; Israel.\\
% ...
$^{a}$ Also at Affiliated with an institute formerly covered by a cooperation agreement with CERN.\\
$^{b}$ Also at An-Najah National University, Nablus; Palestine.\\
$^{c}$ Also at Borough of Manhattan Community College, City University of New York, New York NY; United States of America.\\
$^{d}$ Also at Center for Interdisciplinary Research and Innovation (CIRI-AUTH), Thessaloniki; Greece.\\
$^{e}$ Also at Centre of Physics of the Universities of Minho and Porto (CF-UM-UP); Portugal.\\
$^{f}$ Also at CERN, Geneva; Switzerland.\\
% ...

\end{flushleft}


%% file: ANA-HIGP-2024-21-PAPER.bib
@article{Heinemeyer:2013tqa,
      author                    = "{LHC Higgs Cross Section Working Group}",
      title                     = "{Handbook of LHC Higgs Cross Sections: 3. Higgs Properties}",
      collaboration  = "LHC Higgs Cross Section Working Group",
      journal                   = "CERN-2013-004",
      year                      = "2013",
      eprint                    = "1307.1347",
      archivePrefix     = "arXiv",
      primaryClass      = "hep-ph",
      doi                       = "10.5170/CERN-2013-004",
      SLACcitation      = "%%CITATION = ARXIV:1307.1347;%%",
}

@article{dipoleRecoil,
  author = "{B. Cabouat and T. Sjöstrand}",
  title = "{Some dipole shower studies}",
  journal        = "Eur. Phys. J. C",
  volume         = "78",
  year           = "2018",
  pages          = "226",
  doi            = "10.1140/epjc/s10052-018-5645-z",
  eprint         = "1710.00391",
  archivePrefix  = "arXiv",
  primaryClass   = "hep-ph",
}

@article{paszke2019pytorch,
  title="{PyTorch: An Imperative Style, High-Performance Deep Learning Library}",
  author="{A. Paszke et al.}",
  journal={NeurIPS},
  volume={32},
  year={2019},
  pages = {8024},
  eprint         = "1912.01703",
  archivePrefix  = "arXiv",
  primaryClass   = "cs.LG",
}

@misc{kingma2017adammethodstochasticoptimization,
      title="{Adam: A Method for Stochastic Optimization}", 
      author="{D. Kingma and J. Ba}",
      year={2017},
      eprint={1412.6980},
      archivePrefix={arXiv},
      primaryClass={cs.LG},
}

@Article{CMS-HIG-21-019-own,
    author         = "{CMS Collaboration}",
    title          = "{Measurement of the Higgs boson mass and width using the four-lepton final state in proton--proton collisions at \(\sqrt{s} = 13\,\text{TeV}\)}",
    journal        = "Phys. Rev. D",
    volume         = "111",
    year           = "2025",
    pages          = "092014",
    doi            = "10.1103/PhysRevD.111.092014",
    reportNumber   = "CERN-EP-2024-210",
    eprint         = "2409.13663",
    archivePrefix  = "arXiv",
    primaryClass   = "hep-ex",
}


%% file: ATLAS-useful.bib
@Article{Cacciari:2008gp,
     author    = "Cacciari, Matteo and Salam, Gavin P. and Soyez, Gregory",
     title     = "{The anti-\(k_{t}\) jet clustering algorithm}",
     journal   = "JHEP",
     volume    = "04",
     year      = "2008",
     pages     = "063",
     eprint    = "0802.1189",
     archivePrefix = "arXiv",
     primaryClass  =  "hep-ph",
     doi       = "10.1088/1126-6708/2008/04/063",
     SLACcitation  = "%%CITATION = 0802.1189;%%"
}

@Article{Fastjet,
      author         = "Cacciari, Matteo and Salam, Gavin P. and Soyez, Gregory",
      title          = "{FastJet user manual}",
      journal        = "Eur. Phys. J. C",
      volume         = "72",
      year           = "2012",
      pages          = "1896",
      doi            = "10.1140/epjc/s10052-012-1896-2",
      eprint         = "1111.6097",
      archivePrefix  = "arXiv",
      primaryClass   = "hep-ph",
      reportNumber   = "CERN-PH-TH-2011-297",
      SLACcitation   = "%%CITATION = ARXIV:1111.6097;%%"
}

@article{Ball:2012cx,
      author         = "{NNPDF Collaboration} and Ball, Richard D. and others",
      title          = "{Parton distributions with LHC data}",
      journal        = "Nucl. Phys. B",
      volume         = "867",
      year           = "2013",
      pages          = "244",
      doi            = "10.1016/j.nuclphysb.2012.10.003",
      eprint         = "1207.1303",
      archivePrefix  = "arXiv",
      primaryClass   = "hep-ph",
      reportNumber   = "EDINBURGH-2012-08, IFUM-FT-997, FR-PHENO-2012-014,
                        RWTH-TTK-12-25, CERN-PH-TH-2012-037, SFB-CPP-12-47\,
                        --CERN-PH-TH-2012-037",
      SLACcitation   = "%%CITATION = ARXIV:1207.1303;%%"
}

@article{Ball:2014uwa,
      author         = "{NNPDF Collaboration} and Ball, Richard D. and others",
      title          = "{Parton distributions for the LHC run II}",
      journal        = "JHEP",
      volume         = "04",
      year           = "2015",
      pages          = "040",
      doi            = "10.1007/JHEP04(2015)040",
      eprint         = "1410.8849",
      archivePrefix  = "arXiv",
      primaryClass   = "hep-ph",
      reportNumber   = "EDINBURGH-2014-15, IFUM-1034-FT, CERN-PH-TH-2013-253,
                        OUTP-14-11P, CAVENDISH-HEP-14-11",
      SLACcitation   = "%%CITATION = ARXIV:1410.8849;%%"
}

@article{Sjostrand:2014zea,
      author         = "Sj{\"o}strand, Torbj{\"o}rn and Ask, Stefan and Christiansen,
                        Jesper R. and Corke, Richard and Desai, Nishita and Ilten,
                        Philip and Mrenna, Stephen and Prestel, Stefan and
                        Rasmussen, Christine O. and Skands, Peter Z.",
      title          = "{An introduction to PYTHIA 8.2}",
      journal        = "Comput. Phys. Commun.",
      volume         = "191",
      year           = "2015",
      pages          = "159",
      doi            = "10.1016/j.cpc.2015.01.024",
      eprint         = "1410.3012",
      archivePrefix  = "arXiv",
      primaryClass   = "hep-ph",
      reportNumber   = "LU-TP-14-36, MCNET-14-22, CERN-PH-TH-2014-190,
                        FERMILAB-PUB-14-316-CD, DESY-14-178, SLAC-PUB-16122,
                        --FERMILAB-PUB-14-316-CD",
      SLACcitation   = "%%CITATION = ARXIV:1410.3012;%%"
}

@Article{Lange:2001uf,
      author         = "Lange, D. J.",
      title          = "{The EvtGen particle decay simulation package}",
      booktitle      = "{Proceedings, 7th International Conference on B physics
                        at hadron machines (BEAUTY 2000)}",
      journal        = "Nucl. Instrum. Meth. A",
      volume         = "462",
      year           = "2001",
      pages          = "152",
      doi            = "10.1016/S0168-9002(01)00089-4",
      SLACcitation   = "%%CITATION = NUIMA,A462,152;%%"
}

@Article{Alwall:2014hca,
      author         = "Alwall, J. and Frederix, R. and Frixione, S. and Hirschi,
                        V. and Maltoni, F. and Mattelaer, O. and Shao, H. -S. and
                        Stelzer, T. and Torrielli, P. and Zaro, M.",
      title          = "{The automated computation of tree-level and
                        next-to-leading order differential cross sections, and
                        their matching to parton shower simulations}",
      journal        = "JHEP",
      volume         = "07",
      year           = "2014",
      pages          = "079",
      doi            = "10.1007/JHEP07(2014)079",
      eprint         = "1405.0301",
      archivePrefix  = "arXiv",
      primaryClass   = "hep-ph",
      reportNumber   = "CERN-PH-TH-2014-064, CP3-14-18, LPN14-066, MCNET-14-09,
                        ZU-TH-14-14",
      SLACcitation   = "%%CITATION = ARXIV:1405.0301;%%"
}

@Article{Bahr:2008pv,
      author         = "B{\"a}hr, M. and others",
      title          = "{Herwig++ physics and manual}",
      journal        = "Eur. Phys. J. C",
      volume         = "58",
      year           = "2008",
      pages          = "639",
      doi            = "10.1140/epjc/s10052-008-0798-9",
      eprint         = "0803.0883",
      archivePrefix  = "arXiv",
      primaryClass   = "hep-ph",
      reportNumber   = "CERN-PH-TH-2008-038, CAVENDISH-HEP-08-03, KA-TP-05-2008,
                        DCPT-08-22, IPPP-08-11, CP3-08-05",
      SLACcitation   = "%%CITATION = ARXIV:0803.0883;%%"
}

@Article{Nason:2004rx,
      author         = "Nason, Paolo",
      title          = "{A new method for combining NLO QCD with shower Monte Carlo algorithms}",
      journal        = "JHEP",
      volume         = "11",
      pages          = "040",
      doi            = "10.1088/1126-6708/2004/11/040",
      year           = "2004",
      eprint         = "hep-ph/0409146",
      archivePrefix  = "arXiv",
}

@Article{Frixione:2007vw,
      author         = "Frixione, Stefano and Nason, Paolo and Oleari, Carlo",
      title          = "{Matching NLO QCD computations with parton shower
                        simulations: the POWHEG method}",
      journal        = "JHEP",
      volume         = "11",
      pages          = "070",
      doi            = "10.1088/1126-6708/2007/11/070",
      year           = "2007",
      eprint         = "0709.2092",
      archivePrefix  = "arXiv",
      primaryClass   = "hep-ph",
}

@Article{Alioli:2010xd,
      author         = "Alioli, Simone and Nason, Paolo and Oleari, Carlo and Re,
                        Emanuele",
      title          = "{A general framework for implementing NLO calculations in
                        shower Monte Carlo programs: the POWHEG BOX}",
      journal        = "JHEP",
      volume         = "06",
      pages          = "043",
      doi            = "10.1007/JHEP06(2010)043",
      year           = "2010",
      eprint         = "1002.2581",
      archivePrefix  = "arXiv",
      primaryClass   = "hep-ph",
}

@article{deFlorian:2016spz,
      author         = "de Florian, D. and others",
      title          = "{Handbook of LHC Higgs Cross Sections: 4. Deciphering the Nature of the Higgs Sector}",
      collaboration  = "LHC Higgs Cross Section Working Group",
      doi            = "10.23731/CYRM-2017-002",
      year           = "2017",
      eprint         = "1610.07922",
      archivePrefix  = "arXiv",
      primaryClass   = "hep-ph",
      reportNumber   = "FERMILAB-FN-1025-T, CERN-2017-002-M",
      SLACcitation   = "%%CITATION = ARXIV:1610.07922;%%"
}

@article{Bothmann:2019yzt,
      author = "Bothmann, Enrico and others",
      title = "{Event generation with Sherpa 2.2}",
      journal = "SciPost Phys.",
      volume = "7",
      year = "2019",
      number = "3",
      pages = "034",
      doi = "10.21468/SciPostPhys.7.3.034",
      reportNumber = "FERMILAB-PUB-19-218-T, SLAC-PUB-17433, IPPP/19/42, MCNET-19-11",
      eprint = "1905.09127",
      archivePrefix = "arXiv",
      primaryClass = "hep-ph",
}

@article{Hamilton:2013fea,
      author         = "Hamilton, Keith and Nason, Paolo and Re, Emanuele and
                        Zanderighi, Giulia",
      title          = "{NNLOPS simulation of Higgs boson production}",
      journal        = "JHEP",
      volume         = "10",
      year           = "2013",
      pages          = "222",
      doi            = "10.1007/JHEP10(2013)222",
      eprint         = "1309.0017",
      archivePrefix  = "arXiv",
      primaryClass   = "hep-ph",
      reportNumber   = "MCNET-13-11, CERN-PH-TH-2013-205, OUTP-13-18P",
      SLACcitation   = "%%CITATION = ARXIV:1309.0017;%%"
}

@article{Hamilton:2015nsa,
      author         = "Hamilton, Keith and Nason, Paolo and Zanderighi, Giulia",
      title          = "{Finite quark-mass effects in the NNLOPS POWHEG+MiNLO
                        Higgs generator}",
      journal        = "JHEP",
      volume         = "05",
      year           = "2015",
      pages          = "140",
      doi            = "10.1007/JHEP05(2015)140",
      eprint         = "1501.04637",
      archivePrefix  = "arXiv",
      primaryClass   = "hep-ph",
      reportNumber   = "CERN-PH-TH-2015-006",
      SLACcitation   = "%%CITATION = ARXIV:1501.04637;%%"
}

@article{Campbell:2012am,
      author         = "Campbell, John M. and Ellis, R. Keith and Frederix,
                        Rikkert and Nason, Paolo and Oleari, Carlo and Williams,
                        Ciaran",
      title          = "{NLO Higgs boson production plus one and two jets using
                        the POWHEG BOX, MadGraph4 and MCFM}",
      journal        = "JHEP",
      volume         = "07",
      year           = "2012",
      pages          = "092",
      doi            = "10.1007/JHEP07(2012)092",
      eprint         = "1202.5475",
      archivePrefix  = "arXiv",
      primaryClass   = "hep-ph",
      reportNumber   = "FERMILAB-PUB-12-040-T, CERN-PH-TH-2012-048",
      SLACcitation   = "%%CITATION = ARXIV:1202.5475;%%"
}

@article{Nason:2009ai,
      author         = "Nason, Paolo and Oleari, Carlo",
      title          = "{NLO Higgs boson production via vector-boson fusion
                        matched with shower in POWHEG}",
      journal        = "JHEP",
      volume         = "02",
      year           = "2010",
      pages          = "037",
      doi            = "10.1007/JHEP02(2010)037",
      eprint         = "0911.5299",
      archivePrefix  = "arXiv",
      primaryClass   = "hep-ph",
      SLACcitation   = "%%CITATION = ARXIV:0911.5299;%%"
}

@article{Agostinelli:2002hh,
      author         = "Agostinelli, S. and others",
      title          = "{\textsc{Geant4} -- a simulation toolkit}",
      journal        = "Nucl. Instrum. Meth. A",
      volume         = "506",
      year           = "2003",
      pages          = "250",
      doi            = "10.1016/S0168-9002(03)01368-8",
      reportNumber   = "SLAC-PUB-9350, FERMILAB-PUB-03-339",
      SLACcitation   = "%%CITATION = NUIMA,A506,250;%%"
}

@Article{Read:2002hq,
    Author = {Read, Alexander L.},
    Journal = {J. Phys. G},
    Pages = {2693},
    doi = "10.1088/0954-3899/28/10/313",
    Title = "{Presentation of search results: the \(CL_s\) technique}",
    Volume = {28},
    Year = {2002}
}

@Article{Cowan:2010js,
   Author = {Cowan, Glen and Cranmer, Kyle and Gross, Eilam and Vitells, Ofer},
   Title = "{Asymptotic formulae for likelihood-based tests of new physics}",
   Journal = {Eur. Phys. J. C},
   Volume = {71},
   Year = {2011},
   Pages = {1554},
   doi = "10.1140/epjc/s10052-011-1554-0",
   Eprint = "1007.1727",
   Archiveprefix = {arXiv},
   Primaryclass = {physics.data-an},
   related = "Cowan:2010js-err",
   relatedstring  = "Erratum:",
}


%% file: ATLAS.bib
@Article{PERF-2007-01,
    author         = "{ATLAS Collaboration}",
    title          = "{The ATLAS Experiment at the CERN Large Hadron Collider}",
    journal        = "JINST",
    volume         = "3",
    year           = "2008",
    pages          = "S08003",
    doi            = "10.1088/1748-0221/3/08/S08003",
    primaryClass   = "hep-ex",
}

@Article{SOFT-2010-01,
    author         = "{ATLAS Collaboration}",
    title          = "{The ATLAS Simulation Infrastructure}",
    journal        = "Eur. Phys. J. C",
    volume         = "70",
    year           = "2010",
    pages          = "823",
    doi            = "10.1140/epjc/s10052-010-1429-9",
    eprint         = "1005.4568",
    archivePrefix  = "arXiv",
    primaryClass   = "physics.ins-det",
}

@Article{HIGG-2012-27,
    author         = "{ATLAS Collaboration}",
    title          = "{Observation of a new particle in the search for the Standard Model Higgs boson with the ATLAS detector at the LHC}",
    journal        = "Phys. Lett. B",
    volume         = "716",
    year           = "2012",
    pages          = "1",
    doi            = "10.1016/j.physletb.2012.08.020",
    reportNumber   = "CERN-PH-EP-2012-218",
    eprint         = "1207.7214",
    archivePrefix  = "arXiv",
    primaryClass   = "hep-ex",
}

@Article{HIGG-2013-01,
    author         = "{ATLAS Collaboration}",
    title          = "{Evidence for the spin-0 nature of the Higgs boson using ATLAS data}",
    journal        = "Phys. Lett. B",
    volume         = "726",
    year           = "2013",
    pages          = "120",
    doi            = "10.1016/j.physletb.2013.08.026",
    reportNumber   = "CERN-PH-EP-2013-102",
    eprint         = "1307.1432",
    archivePrefix  = "arXiv",
    primaryClass   = "hep-ex",
}

@Article{HIGG-2013-12,
    author         = "{ATLAS Collaboration}",
    title          = "{Measurement of the Higgs boson mass from the \(H\rightarrow \gamma\gamma\) and \(H \rightarrow ZZ^{*} \rightarrow 4\ell\) channels in \(pp\) collisions at center-of-mass energies of \(7\) and \(8\,\text{TeV}\) with the ATLAS detector}",
    journal        = "Phys. Rev. D",
    volume         = "90",
    year           = "2014",
    pages          = "052004",
    doi            = "10.1103/PhysRevD.90.052004",
    reportNumber   = "CERN-PH-EP-2014-122",
    eprint         = "1406.3827",
    archivePrefix  = "arXiv",
    primaryClass   = "hep-ex",
}

@Article{HIGG-2013-13,
    author         = "{ATLAS Collaboration}",
    title          = "{Observation and measurement of Higgs boson decays to \(WW^*\) with the ATLAS detector}",
    journal        = "Phys. Rev. D",
    volume         = "92",
    year           = "2015",
    pages          = "012006",
    doi            = "10.1103/PhysRevD.92.012006",
    reportNumber   = "CERN-PH-EP-2014-270",
    eprint         = "1412.2641",
    archivePrefix  = "arXiv",
    primaryClass   = "hep-ex",
}

@Article{HIGG-2013-14,
    author         = "{ATLAS Collaboration}",
    title          = "{Determination of spin and parity of the Higgs boson in the \(WW^*\rightarrow e \nu \mu \nu\) decay channel with the ATLAS detector}",
    journal        = "Eur. Phys. J. C",
    volume         = "75",
    year           = "2015",
    pages          = "231",
    doi            = "10.1140/epjc/s10052-015-3436-3",
    reportNumber   = "CERN-PH-EP-2015-037",
    eprint         = "1503.03643",
    archivePrefix  = "arXiv",
    primaryClass   = "hep-ex",
}

@Article{HIGG-2013-15,
    author         = "{ATLAS Collaboration}",
    title          = "{Study of \((W/Z)H\) production and Higgs boson couplings using \(H \rightarrow WW^{\ast}\) decays with the ATLAS detector}",
    journal        = "JHEP",
    volume         = "08",
    year           = "2015",
    pages          = "137",
    doi            = "10.1007/JHEP08(2015)137",
    reportNumber   = "CERN-PH-EP-2015-104",
    eprint         = "1506.06641",
    archivePrefix  = "arXiv",
    primaryClass   = "hep-ex",
}

@Article{HIGG-2013-17,
    author         = "{ATLAS Collaboration}",
    title          = "{Study of the spin and parity of the Higgs boson in diboson decays with the ATLAS detector}",
    journal        = "Eur. Phys. J. C",
    volume         = "75",
    year           = "2015",
    pages          = "476",
    doi            = "10.1140/epjc/s10052-015-3685-1",
    reportNumber   = "CERN-PH-EP-2015-114",
    eprint         = "1506.05669",
    archivePrefix  = "arXiv",
    primaryClass   = "hep-ex",
    related        = "HIGG-2013-17-err",
    relatedstring  = "Erratum:",
}

@Article{HIGG-2013-32,
    author         = "{ATLAS Collaboration}",
    title          = "{Evidence for the Higgs-boson Yukawa coupling to tau leptons with the ATLAS detector}",
    journal        = "JHEP",
    volume         = "04",
    year           = "2015",
    pages          = "117",
    doi            = "10.1007/JHEP04(2015)117",
    reportNumber   = "CERN-PH-EP-2014-262",
    eprint         = "1501.04943",
    archivePrefix  = "arXiv",
    primaryClass   = "hep-ex",
}

@Article{HIGG-2014-14,
    author         = "{ATLAS and CMS Collaborations}",
    title          = "{Combined Measurement of the Higgs Boson Mass in \(pp\) Collisions at \(\sqrt{s} = 7\) and \(8\,\text{TeV}\) with the ATLAS and CMS Experiments}",
    journal        = "Phys. Rev. Lett.",
    volume         = "114",
    year           = "2015",
    pages          = "191803",
    doi            = "10.1103/PhysRevLett.114.191803",
    reportNumber   = "CERN-PH-EP-2015-075",
    eprint         = "1503.07589",
    archivePrefix  = "arXiv",
    primaryClass   = "hep-ex",
}

@Article{PERF-2015-09,
    author         = "{ATLAS Collaboration}",
    title          = "{Jet reconstruction and performance using particle flow with the ATLAS Detector}",
    journal        = "Eur. Phys. J. C",
    volume         = "77",
    year           = "2017",
    pages          = "466",
    doi            = "10.1140/epjc/s10052-017-5031-2",
    reportNumber   = "CERN-EP-2017-024",
    eprint         = "1703.10485",
    archivePrefix  = "arXiv",
    primaryClass   = "hep-ex",
}

@Article{HIGG-2016-22,
    author         = "{ATLAS Collaboration}",
    title          = "{Measurement of the Higgs boson coupling properties in the \(H \rightarrow ZZ^{*} \rightarrow 4\ell\) decay channel at \(\sqrt{s} = 13\,\text{TeV}\) with the ATLAS detector}",
    journal        = "JHEP",
    volume         = "03",
    year           = "2018",
    pages          = "095",
    doi            = "10.1007/JHEP03(2018)095",
    reportNumber   = "CERN-EP-2017-206",
    eprint         = "1712.02304",
    archivePrefix  = "arXiv",
    primaryClass   = "hep-ex",
}

@Article{HIGG-2016-25,
    author         = "{ATLAS Collaboration}",
    title          = "{Measurement of inclusive and differential cross sections in the \(H \rightarrow ZZ^* \rightarrow 4\ell\) decay channel in \(pp\) collisions at \(\sqrt{s} = 13\,\text{TeV}\) with the ATLAS detector}",
    journal        = "JHEP",
    volume         = "10",
    year           = "2017",
    pages          = "132",
    doi            = "10.1007/JHEP10(2017)132",
    reportNumber   = "CERN-EP-2017-139",
    eprint         = "1708.02810",
    archivePrefix  = "arXiv",
    primaryClass   = "hep-ex",
}

@Article{HIGG-2016-33,
    author         = "{ATLAS Collaboration}",
    title          = "{Measurement of the Higgs boson mass in the \(H \rightarrow ZZ^* \rightarrow 4\ell\) and \(H \rightarrow \gamma\gamma\) channels with \(\sqrt{s} = 13\,\text{TeV}\) \(pp\) collisions using the ATLAS detector}",
    journal        = "Phys. Lett. B",
    volume         = "784",
    year           = "2018",
    pages          = "345",
    doi            = "10.1016/j.physletb.2018.07.050",
    reportNumber   = "CERN-EP-2018-085",
    eprint         = "1806.00242",
    archivePrefix  = "arXiv",
    primaryClass   = "hep-ex",
}

@Article{HIGG-2017-11,
    author         = "{ATLAS Collaboration}",
    title          = "{Combined measurement of differential and total cross sections in the \(H \rightarrow \gamma \gamma\) and the \(H \rightarrow ZZ^* \rightarrow 4\ell\) decay channels at \(\sqrt{s} = 13\,\text{TeV}\) with the ATLAS detector}",
    journal        = "Phys. Lett. B",
    volume         = "786",
    year           = "2018",
    pages          = "114",
    doi            = "10.1016/j.physletb.2018.09.019",
    reportNumber   = "CERN-EP-2018-080",
    eprint         = "1805.10197",
    archivePrefix  = "arXiv",
    primaryClass   = "hep-ex",
}

@Article{FTAG-2018-01,
    author         = "{ATLAS Collaboration}",
    title          = "{ATLAS \(b\)-jet identification performance and efficiency measurement with \(t\bar{t}\) events in \(pp\) collisions at \(\sqrt{s} = 13\,\text{TeV}\)}",
    journal        = "Eur. Phys. J. C",
    volume         = "79",
    year           = "2019",
    pages          = "970",
    doi            = "10.1140/epjc/s10052-019-7450-8",
    reportNumber   = "CERN-EP-2019-132",
    eprint         = "1907.05120",
    archivePrefix  = "arXiv",
    primaryClass   = "hep-ex",
}

@Article{HIGG-2018-04,
    author         = "{ATLAS Collaboration}",
    title          = "{Observation of \(H \rightarrow b\bar{b}\) decays and \(VH\) production with the ATLAS detector}",
    journal        = "Phys. Lett. B",
    volume         = "786",
    year           = "2018",
    pages          = "59",
    doi            = "10.1016/j.physletb.2018.09.013",
    reportNumber   = "CERN-EP-2018-215",
    eprint         = "1808.08238",
    archivePrefix  = "arXiv",
    primaryClass   = "hep-ex",
}

@Article{JETM-2018-05,
    author         = "{ATLAS Collaboration}",
    title          = "{Jet energy scale and resolution measured in proton--proton collisions at \(\sqrt{s} = 13\,\text{TeV}\) with the ATLAS detector}",
    journal        = "Eur. Phys. J. C",
    volume         = "81",
    year           = "2021",
    pages          = "689",
    doi            = "10.1140/epjc/s10052-021-09402-3",
    reportNumber   = "CERN-EP-2020-083",
    eprint         = "2007.02645",
    archivePrefix  = "arXiv",
    primaryClass   = "hep-ex",
}

@Article{MUON-2018-03,
    author         = "{ATLAS Collaboration}",
    title          = "{Muon reconstruction and identification efficiency in ATLAS using the full Run~2 \(pp\) collision data set at \(\sqrt{s} = 13\,\text{TeV}\)}",
    journal        = "Eur. Phys. J. C",
    volume         = "81",
    year           = "2021",
    pages          = "578",
    doi            = "10.1140/epjc/s10052-021-09233-2",
    reportNumber   = "CERN-EP-2020-199",
    eprint         = "2012.00578",
    archivePrefix  = "arXiv",
    primaryClass   = "hep-ex",
}

@Article{TRIG-2018-01,
    author         = "{ATLAS Collaboration}",
    title          = "{Performance of the ATLAS muon triggers in Run~2}",
    journal        = "JINST",
    volume         = "15",
    year           = "2020",
    pages          = "P09015",
    doi            = "10.1088/1748-0221/15/09/p09015",
    reportNumber   = "CERN-EP-2020-031",
    eprint         = "2004.13447",
    archivePrefix  = "arXiv",
    primaryClass   = "physics.ins-det",
}

@Article{FTAG-2019-02,
    author         = "{ATLAS Collaboration}",
    title          = "{Calibration of the light-flavour jet mistagging efficiency of the \(b\)-tagging algorithms with \(Z\)+jets events using \(139\,\text{fb}^{-1}\) of ATLAS proton--proton collision data at \(\sqrt{s} = 13\,\text{TeV}\)}",
    journal        = "Eur. Phys. J. C",
    volume         = "83",
    year           = "2023",
    pages          = "728",
    doi            = "10.1140/epjc/s10052-023-11736-z",
    reportNumber   = "CERN-EP-2022-211",
    eprint         = "2301.06319",
    archivePrefix  = "arXiv",
    primaryClass   = "hep-ex",
}

@Article{FTAG-2019-07,
    author         = "{ATLAS Collaboration}",
    title          = "{ATLAS flavour-tagging algorithms for the LHC Run~2 \(pp\) collision dataset}",
    journal        = "Eur. Phys. J. C",
    volume         = "83",
    year           = "2023",
    pages          = "681",
    doi            = "10.1140/epjc/s10052-023-11699-1",
    reportNumber   = "CERN-EP-2022-226",
    eprint         = "2211.16345",
    archivePrefix  = "arXiv",
    primaryClass   = "physics.data-an",
}

@Article{HIGG-2019-04,
    author         = "{ATLAS Collaboration}",
    title          = "{Measurements of Higgs bosons decaying to bottom quarks from vector boson fusion production with the ATLAS experiment at \(\sqrt{s} = 13\,\text{TeV}\)}",
    journal        = "Eur. Phys. J. C",
    volume         = "81",
    year           = "2021",
    pages          = "537",
    doi            = "10.1140/epjc/s10052-021-09192-8",
    reportNumber   = "CERN-EP-2020-195",
    eprint         = "2011.08280",
    archivePrefix  = "arXiv",
    primaryClass   = "hep-ex",
}

@Article{HIGG-2019-16,
    author         = "{ATLAS Collaboration}",
    title          = "{Measurement of the Higgs boson mass with \(H \to \gamma\gamma\) decays in \(140\,\text{fb}^{-1}\) of \(\sqrt{s} = 13\,\text{TeV}\) \(pp\) collisions with the ATLAS detector}",
    journal        = "Phys. Lett. B",
    volume         = "847",
    year           = "2023",
    pages          = "138315",
    doi            = "10.1016/j.physletb.2023.138315",
    reportNumber   = "CERN-EP-2023-160",
    eprint         = "2308.07216",
    archivePrefix  = "arXiv",
    primaryClass   = "hep-ex",
}

@Article{TRIG-2019-04,
    author         = "{ATLAS Collaboration}",
    title          = "{Operation of the ATLAS trigger system in Run~2}",
    journal        = "JINST",
    volume         = "15",
    year           = "2020",
    pages          = "P10004",
    doi            = "10.1088/1748-0221/15/10/P10004",
    reportNumber   = "CERN-EP-2020-109",
    eprint         = "2007.12539",
    archivePrefix  = "arXiv",
    primaryClass   = "physics.ins-det",
}

@Article{FTAG-2020-08,
    author         = "{ATLAS Collaboration}",
    title          = "{Measurement of the \(c\)-jet mistagging efficiency in \(t\bar{t}\) events using \(pp\) collision data at \(\sqrt{s} = 13\,\text{TeV}\) collected with the ATLAS detector}",
    journal        = "Eur. Phys. J. C",
    volume         = "82",
    year           = "2022",
    pages          = "95",
    doi            = "10.1140/epjc/s10052-021-09843-w",
    reportNumber   = "CERN-EP-2021-166",
    eprint         = "2109.10627",
    archivePrefix  = "arXiv",
    primaryClass   = "hep-ex",
}

@Article{HIGG-2020-07,
    author         = "{ATLAS Collaboration}",
    title          = "{Measurement of the Higgs boson mass in the \(H \to ZZ^* \to 4\ell\) decay channel using \(139\,\text{fb}^{-1}\) of \(\sqrt{s} = 13\,\text{TeV}\) \(pp\) collisions recorded by the ATLAS detector at the LHC}",
    journal        = "Phys. Lett. B",
    volume         = "843",
    year           = "2023",
    pages          = "137880",
    doi            = "10.1016/j.physletb.2023.137880",
    reportNumber   = "CERN-EP-2022-135",
    eprint         = "2207.00320",
    archivePrefix  = "arXiv",
    primaryClass   = "hep-ex",
}

@Article{HIGG-2020-20,
    author         = "{ATLAS Collaboration}",
    title          = "{Measurements of \(WH\) and \(ZH\) production with Higgs boson decays into bottom quarks and direct constraints on the charm Yukawa coupling in \(13\,\text{TeV}\) \(pp\) collisions with the ATLAS detector}",
    journal        = "JHEP",
    volume         = "04",
    year           = "2025",
    pages          = "075",
    doi            = "10.1007/JHEP04(2025)075",
    reportNumber   = "CERN-EP-2024-237",
    eprint         = "2410.19611",
    archivePrefix  = "arXiv",
    primaryClass   = "hep-ex",
}

@Article{HIGG-2020-24,
    author         = "{ATLAS Collaboration}",
    title          = "{Measurement of the associated production of a top-antitop-quark pair and a Higgs boson decaying into a \(b\bar{b}\) pair in \(pp\) collisions at \(\sqrt{s} = 13\,\text{TeV}\) using the ATLAS detector at the LHC}",
    journal        = "Eur. Phys. J. C",
    volume         = "85",
    year           = "2025",
    pages          = "210",
    doi            = "10.1140/epjc/s10052-025-13740-x",
    reportNumber   = "CERN-EP-2024-194",
    eprint         = "2407.10904",
    archivePrefix  = "arXiv",
    primaryClass   = "hep-ex",
}

@Article{DAPR-2021-01,
    author         = "{ATLAS Collaboration}",
    title          = "{Luminosity determination in \(pp\) collisions at \(\sqrt{s} = 13\,\text{TeV}\) using the ATLAS detector at the LHC}",
    journal        = "Eur. Phys. J. C",
    volume         = "83",
    year           = "2023",
    pages          = "982",
    doi            = "10.1140/epjc/s10052-023-11747-w",
    reportNumber   = "CERN-EP-2022-281",
    eprint         = "2212.09379",
    archivePrefix  = "arXiv",
    primaryClass   = "hep-ex",
}

@Article{EGAM-2021-01,
    author         = "{ATLAS Collaboration}",
    title          = "{Electron and photon efficiencies in LHC Run~2 with the ATLAS experiment}",
    journal        = "JHEP",
    volume         = "05",
    year           = "2024",
    pages          = "162",
    doi            = "10.1007/JHEP05(2024)162",
    reportNumber   = "CERN-EP-2023-182",
    eprint         = "2308.13362",
    archivePrefix  = "arXiv",
    primaryClass   = "hep-ex",
}

@Article{HIGG-2022-04,
    author         = "{ATLAS Collaboration}",
    title          = "{Measurement of the total and differential Higgs boson production cross-sections at \(\sqrt{s} = 13\,\text{TeV}\) with the ATLAS detector by combining the \(H \to ZZ^* \to 4\ell\) and \(H \to \gamma \gamma\) decay channels}",
    journal        = "JHEP",
    volume         = "05",
    year           = "2023",
    pages          = "028",
    doi            = "10.1007/JHEP05(2023)028",
    reportNumber   = "CERN-EP-2022-143",
    eprint         = "2207.08615",
    archivePrefix  = "arXiv",
    primaryClass   = "hep-ex",
}

@Article{SOFT-2022-02,
    author         = "{ATLAS Collaboration}",
    title          = "{Software and computing for Run~3 of the ATLAS experiment at the LHC}",
    journal        = "Eur. Phys. J. C",
    volume         = "85",
    year           = "2025",
    pages          = "234",
    doi            = "10.1140/epjc/s10052-024-13701-w",
    reportNumber   = "CERN-EP-2024-100",
    eprint         = "2404.06335",
    archivePrefix  = "arXiv",
    primaryClass   = "hep-ex",
}

@Article{TRIG-2022-01,
    author         = "{ATLAS Collaboration}",
    title          = "{The ATLAS trigger system for LHC Run~3 and trigger performance in 2022}",
    journal        = "JINST",
    volume         = "19",
    year           = "2024",
    pages          = "P06029",
    doi            = "10.1088/1748-0221/19/06/P06029",
    reportNumber   = "CERN-EP-2023-299",
    eprint         = "2401.06630",
    archivePrefix  = "arXiv",
    primaryClass   = "hep-ex",
}

@Article{FTAG-2023-05,
    author         = "{ATLAS Collaboration}",
    title          = "{Transforming jet flavour tagging at ATLAS}",
    year           = "2025",
    reportNumber   = "CERN-EP-2025-103",
    eprint         = "2505.19689",
    archivePrefix  = "arXiv",
    primaryClass   = "hep-ex",
}

@Article{HIGP-2024-04,
    author         = "{ATLAS Collaboration}",
    title          = "{Search for Higgs bosons produced in association with a high-energy photon via vector-boson fusion and decaying to a pair of \(b\)-quarks in the ATLAS detector}",
    year           = "2025",
    reportNumber   = "CERN-EP-2025-110",
    eprint         = "2509.14005",
    archivePrefix  = "arXiv",
    primaryClass   = "hep-ex",
}

@Article{HIGP-2024-11,
    author         = "{ATLAS Collaboration}",
    title          = "{Evidence for the dimuon decay of the Higgs boson in \(pp\) collisions with the ATLAS detector}",
    year           = "2025",
    reportNumber   = "CERN-EP-2025-149",
    eprint         = "2507.03595",
    archivePrefix  = "arXiv",
    primaryClass   = "hep-ex",
}

@Booklet{ATL-SOFT-PUB-2025-001,
    author         = "{ATLAS Collaboration}",
    title          = "{ATLAS Computing Acknowledgements}",
    howpublished   = "{ATL-SOFT-PUB-2025-001}",
    url            = "https://cds.cern.ch/record/2922210",
    year           = "2025",
}


%% file: CMS.bib
@Article{CMS-HIG-12-028,
    author         = "{CMS Collaboration}",
    title          = "{Observation of a new boson at a mass of 125 GeV with the CMS experiment at the LHC}",
    journal        = "Phys. Lett. B",
    volume         = "716",
    year           = "2012",
    pages          = "30",
    doi            = "10.1016/j.physletb.2012.08.021",
    reportNumber   = "CERN-PH-EP-2012-220",
    eprint         = "1207.7235",
    archivePrefix  = "arXiv",
    primaryClass   = "hep-ex",
}

@Article{CMS-HIG-12-041,
    author         = "{CMS Collaboration}",
    title          = "{Study of the Mass and Spin-Parity of the Higgs Boson Candidate Via Its Decays to \(Z\) Boson Pairs}",
    journal        = "Phys. Rev. Lett.",
    volume         = "110",
    year           = "2013",
    pages          = "081803",
    doi            = "10.1103/PhysRevLett.110.081803",
    reportNumber   = "CERN-PH-EP-2012-372",
    eprint         = "1212.6639",
    archivePrefix  = "arXiv",
    primaryClass   = "hep-ex",
}

@Article{CMS-HIG-13-002,
    author         = "{CMS Collaboration}",
    title          = "{Measurement of the properties of a Higgs boson in the four-lepton final state}",
    journal        = "Phys. Rev. D",
    volume         = "89",
    year           = "2014",
    pages          = "092007",
    doi            = "10.1103/PhysRevD.89.092007",
    reportNumber   = "CERN-PH-EP-2013-220",
    eprint         = "1312.5353",
    archivePrefix  = "arXiv",
    primaryClass   = "hep-ex",
}

@Article{CMS-HIG-13-023,
    author         = "{CMS Collaboration}",
    title          = "{Measurement of Higgs boson production and properties in the \(WW\) decay channel with leptonic final states}",
    journal        = "JHEP",
    volume         = "01",
    year           = "2014",
    pages          = "096",
    doi            = "10.1007/JHEP01(2014)096",
    reportNumber   = "CERN-PH-EP-2013-221",
    eprint         = "1312.1129",
    archivePrefix  = "arXiv",
    primaryClass   = "hep-ex",
}

@Article{CMS-HIG-14-018,
    author         = "{CMS Collaboration}",
    title          = "{Constraints on the spin-parity and anomalous HVV couplings of the Higgs boson in proton collisions at \(7\) and \(8\,\text{TeV}\)}",
    journal        = "Phys. Rev. D",
    volume         = "92",
    year           = "2015",
    pages          = "012004",
    doi            = "10.1103/PhysRevD.92.012004",
    reportNumber   = "CERN-PH-EP-2014-265",
    eprint         = "1411.3441",
    archivePrefix  = "arXiv",
    primaryClass   = "hep-ex",
}

@Article{CMS-HIG-14-028,
    author         = "{CMS Collaboration}",
    title          = "{Measurement of differential and integrated fiducial cross sections for Higgs boson production in the four-lepton decay channel in \(pp\) collisions at \(\sqrt{s} = 7\) and \(8\,\text{TeV}\)}",
    journal        = "JHEP",
    volume         = "04",
    year           = "2016",
    pages          = "005",
    doi            = "10.1007/JHEP04(2016)005",
    reportNumber   = "CERN-PH-EP-2015-285",
    eprint         = "1512.08377",
    archivePrefix  = "arXiv",
    primaryClass   = "hep-ex",
}

@Article{CMS-HIG-16-043,
    author         = "{CMS Collaboration}",
    title          = "{Observation of the Higgs boson decay to a pair of \(\tau\) leptons with the CMS detector}",
    journal        = "Phys. Lett. B",
    volume         = "779",
    year           = "2018",
    pages          = "283",
    doi            = "10.1016/j.physletb.2018.02.004",
    reportNumber   = "CERN-EP-2017-181",
    eprint         = "1708.00373",
    archivePrefix  = "arXiv",
    primaryClass   = "hep-ex",
}

@Article{CMS-HIG-16-044,
    author         = "{CMS Collaboration}",
    title          = "{Evidence for the Higgs boson decay to a bottom quark--antiquark pair}",
    journal        = "Phys. Lett. B",
    volume         = "780",
    year           = "2018",
    pages          = "501",
    doi            = "10.1016/j.physletb.2018.02.050",
    reportNumber   = "CERN-EP-2017-233",
    eprint         = "1709.07497",
    archivePrefix  = "arXiv",
    primaryClass   = "hep-ex",
}

@Article{CMS-HIG-18-002,
    author         = "{CMS Collaboration}",
    title          = "{Measurements of the Higgs boson width and anomalous \(HVV\) couplings from on-shell and off-shell production in the four-lepton final state}",
    journal        = "Phys. Rev. D",
    volume         = "99",
    year           = "2019",
    pages          = "112003",
    doi            = "10.1103/PhysRevD.99.112003",
    reportNumber   = "CERN-EP-2018-329",
    eprint         = "1901.00174",
    archivePrefix  = "arXiv",
    primaryClass   = "hep-ex",
}

@Article{CMS-HIG-18-016,
    author         = "{CMS Collaboration}",
    title          = "{Observation of Higgs Boson Decay to Bottom Quarks}",
    journal        = "Phys. Rev. Lett.",
    volume         = "121",
    year           = "2018",
    pages          = "121801",
    doi            = "10.1103/PhysRevLett.121.121801",
    reportNumber   = "CERN-EP-2018-223",
    eprint         = "1808.08242",
    archivePrefix  = "arXiv",
    primaryClass   = "hep-ex",
}

@Article{CMS-HIG-19-002,
    author         = "{CMS Collaboration}",
    title          = "{Measurement of the inclusive and differential Higgs boson production cross sections in the leptonic \(WW\) decay mode at \(\sqrt{s} = 13\,\text{TeV}\)}",
    journal        = "JHEP",
    volume         = "03",
    year           = "2021",
    pages          = "003",
    doi            = "10.1007/JHEP03(2021)003",
    reportNumber   = "CERN-EP-2020-106",
    eprint         = "2007.01984",
    archivePrefix  = "arXiv",
    primaryClass   = "hep-ex",
}

@Article{CMS-HIG-19-004,
    author         = "{CMS Collaboration}",
    title          = "{A measurement of the Higgs boson mass in the diphoton decay channel}",
    journal        = "Phys. Lett. B",
    volume         = "805",
    year           = "2020",
    pages          = "135425",
    doi            = "10.1016/j.physletb.2020.135425",
    reportNumber   = "CERN-EP-2020-004",
    eprint         = "2002.06398",
    archivePrefix  = "arXiv",
    primaryClass   = "hep-ex",
}

@Article{CMS-HIG-19-006,
    author         = "{CMS Collaboration}",
    title          = "{Evidence for Higgs boson decay to a pair of muons}",
    journal        = "JHEP",
    volume         = "01",
    year           = "2021",
    pages          = "148",
    doi            = "10.1007/JHEP01(2021)148",
    reportNumber   = "CERN-EP-2020-164",
    eprint         = "2009.04363",
    archivePrefix  = "arXiv",
    primaryClass   = "hep-ex",
}

@Article{CMS-HIG-19-011,
    author         = "{CMS Collaboration}",
    title          = "{Measurement of the \(t\bar{t}H\) and \(tH\) production rates in the \(H \rightarrow b\bar{b}\) decay channel using proton--proton collision data at \(\sqrt{s} = 13\,\text{TeV}\)}",
    journal        = "JHEP",
    volume         = "02",
    year           = "2025",
    pages          = "097",
    doi            = "10.1007/JHEP02(2025)097",
    reportNumber   = "CERN-EP-2024-179",
    eprint         = "2407.10896",
    archivePrefix  = "arXiv",
    primaryClass   = "hep-ex",
}

@Article{CMS-HIG-21-008,
    author         = "{CMS Collaboration}",
    title          = "{Search for Higgs Boson Decay to a Charm Quark-Antiquark Pair in Proton--Proton Collisions at \(\sqrt{s} = 13\,\text{TeV}\)}",
    journal        = "Phys. Rev. Lett.",
    volume         = "131",
    year           = "2023",
    pages          = "061801",
    doi            = "10.1103/PhysRevLett.131.061801",
    reportNumber   = "CERN-EP-2022-081",
    eprint         = "2205.05550",
    archivePrefix  = "arXiv",
    primaryClass   = "hep-ex",
}

@Article{CMS-HIG-22-008,
    author         = "{CMS Collaboration}",
    title          = "{Constraints on anomalous Higgs boson couplings from its production and decay using the \(WW\) channel in proton--proton collisions at \(\sqrt{s} = 13\,\text{TeV}\)}",
    journal        = "Eur. Phys. J. C",
    volume         = "84",
    year           = "2024",
    pages          = "779",
    doi            = "10.1140/epjc/s10052-024-12925-0",
    reportNumber   = "CERN-EP-2024-035",
    eprint         = "2403.00657",
    archivePrefix  = "arXiv",
    primaryClass   = "hep-ex",
}

@Article{CMS-HIG-24-018,
    author         = "{CMS Collaboration}",
    title          = "{Simultaneous probe of the charm and bottom quark Yukawa couplings using \(t\bar{t}H\) events}",
    year           = "2025",
    reportNumber   = "CERN-EP-2025-202",
    eprint         = "2509.22535",
    archivePrefix  = "arXiv",
    primaryClass   = "hep-ex",
}


%% file: ConfNotes.bib
@Booklet{ATLAS-CONF-2014-018,
    author         = "{ATLAS Collaboration}",
    title          = "{Tagging and suppression of pileup jets with the ATLAS detector}",
    howpublished   = "{ATLAS-CONF-2014-018}",
    url            = "https://cds.cern.ch/record/1700870",
    year           = "2014",
}


%% file: PubNotes.bib
@Booklet{ATL-PHYS-PUB-2014-021,
    author         = "{ATLAS Collaboration}",
    title          = "{ATLAS Pythia~8 tunes to \(7~\text{TeV}\) data}",
    howpublished   = "{ATL-PHYS-PUB-2014-021}",
    url            = "https://cds.cern.ch/record/1966419",
    year           = "2014",
}

@Booklet{ATL-PHYS-PUB-2019-026,
    author         = "{ATLAS Collaboration}",
    title          = "{Forward jet vertex tagging using the particle flow algorithm}",
    howpublished   = "{ATL-PHYS-PUB-2019-026}",
    url            = "https://cds.cern.ch/record/2683100",
    year           = "2019",
}
